\documentclass[prc,twocolumn,floatfix,superscriptaddress,showpacs]{revtex4}
\usepackage{graphicx,dcolumn,bm}
\usepackage{epstopdf}
\usepackage{enumerate}
\usepackage{pstricks}
\usepackage{dcolumn,amsmath,amssymb}

\begin{document}

\title{Search for a T-odd, P-even Triple Correlation in Neutron Decay}

\author{T.E. Chupp}
 \affiliation{University of Michigan, Ann Arbor, Michigan 48104, USA}

\author{R.L. Cooper}
\affiliation{University of Michigan, Ann Arbor, Michigan 48104, USA}

\author{K.P. Coulter}
\affiliation{University of Michigan, Ann Arbor, Michigan 48104, USA}

\author{S.J. Freedman}
 \affiliation{Physics
Department, University of California, Berkeley, and Lawrence Berkeley
National Laboratory, Berkeley, California 94720, USA}

\author{B.K. Fujikawa}
\affiliation{Physics
Department, University of California, Berkeley, and Lawrence Berkeley
National Laboratory, Berkeley, California 94720, USA}

\author{A. Garc\'{\i}a}
\affiliation{CENPA and Physics Department, University of Washington, Seattle, WA 98195 USA}
\affiliation{Department of Physics, University of Notre Dame, Notre Dame, IN 46556 USA} 

\author{G.L. Jones}
\affiliation{Physics Department, Hamilton College, Clinton, NY 13323, USA}

\author{H.P. Mumm}
\affiliation{National Institute of Standards and Technology, Gaithersburg, MD 20899, USA}
   
\author{J.S. Nico}
\affiliation{National Institute of Standards and Technology, Gaithersburg, MD 20899, USA}

\author{A.K. Thompson}
 \affiliation{National Institute of Standards and Technology, Gaithersburg, MD 20899, USA}
 
\author{C.A. Trull}
\affiliation{Physics Department, Tulane University, New Orleans, LA 70118, USA}

\author{F.E. Wietfeldt}
\affiliation{Physics Department, Tulane University, New Orleans, LA 70118, USA}
  \
   \author{J.F. Wilkerson}
\affiliation{CENPA and Physics Department, University of Washington, Seattle, WA 98195 USA}
 \affiliation{Department of Physics and Astronomy, University of North Carolina, Chapel Hill, North Carolina 27599, USA}
\affiliation{ Oak Ridge National  Lab,  Oak Ridge, TN, 37831 USA}

\let\oldhat\hat
\renewcommand{\vec}[1]{\mathbf{#1}}
\renewcommand{\hat}[1]{\oldhat{\mathbf{#1}}}


\begin{abstract}

{\bf Background:} Time-reversal-invariance violation, or equivalently CP violation, may explain the observed cosmological baryon asymmetry as well as signal physics beyond the Standard Model. In the decay of polarized neutrons, the triple correlation  $D\langle \vec J_n\rangle\cdot(\vec p_e\times \vec p_\nu)$ is a parity-even, time-reversal-odd observable that is uniquely sensitive to the relative phase of the axial-vector amplitude with respect to the vector amplitude. The triple correlation is also sensitive to  possible contributions from scalar and tensor amplitudes. 
Final-state effects also contribute to $D$ at the level of 10$^{-5}$ and can be calculated with a precision of 1\% or better. 
 {\bf Purpose:}  We have improved the sensitivity to T-odd, P-even interactions in nuclear beta decay.  {\bf Methods}: We measured proton-electron coincidences from decays of longitudinally polarized neutrons with a highly symmetric detector array designed to cancel the time-reversal-even, parity-odd Standard-Model contributions to polarized neutron decay.
Over 300 million proton-electron coincidence events were used to extract $D$ and study systematic effects in a blind analysis. 
{\bf Results:}  We find $D=[-0.94\pm 1.89 ({\rm stat})\pm 0.97({\rm sys})]\times 10^{-4}$. {\bf Conclusions:} This is the most sensitive measurement of $D$ in nuclear beta decay. Our result can be interpreted as a measurement of the phase of the ratio of the axial-vector and vector coupling constants ($C_A/C_V=| \lambda |e^{i\phi_{AV}}$ ) with $\phi_{AV} = 180.012^\circ \pm 0.028^\circ$ (68\% confidence level) or to constrain time-reversal violating scalar and tensor interactions that arise in certain extensions to the Standard Model such as leptoquarks. This paper presents details of the experiment, analysis, and systematic-error corrections.
\end{abstract}

\maketitle

\section{Introduction}

The symmetries of physical processes under the transformations of charge conjugation (C), parity  (P), and time reversal (T) have played a central role in the development of the Standard Model of elementary-particle interactions~\cite{rf:Dubbers2011}. Time-reversal-symmetry violation (or T violation), which is equivalent to CP violation assuming CPT symmetry, has been of particular interest because it is sensitive to many kinds of new physics. 
The CP-violating parameters of the Standard Model are the Cabibbo-Kobayashi-Maskawa (CKM) phase, which enters in the mixing of three generations of  quarks, and the parameter $\theta_{QCD}$. The effect of the CKM phase is strongly suppressed in the permanent electric dipole moments (EDMs) of the neutron~\cite{rf:BakerNeutronEDM} and heavy atoms~\cite{rf:Griffith199Hg,rf:Rosenberry2001}, and recent EDM results combine to set upper limits on $\theta_{QCD}$.  
All laboratory measurements  to date are consistent with a single source of CP violation, {\it i.e.}
the phase in the CKM matrix. An exception may be 
the 3.2 sigma deviation observed recently as an asymmetry
in the production of pairs of like-sign muons reported by the D0 collaboration~\cite{rf:D0dimuon}. 

In spite of this success, laboratory and astrophysical observations, which
include neutrinos with non-zero masses, the abundance of non-baryonic dark matter,
and the baryon asymmetry of the universe, provide strong evidence that the Standard Model is incomplete. Generation of the baryon asymmetry requires CP violation that  cannot be accounted for by Standard-Model physics~\cite{rf:Trodden,rf:Davidson2008}. This provides strong motivation to search for new sources of CP violation. Such CP-violating phases would also, in general, affect T-odd observables in neutron decay, in particular, the T-odd/P-even triple correlation in polarized neutron decay. 

We have measured the triple-correlation  in the decay of polarized neutrons at the NIST Center for Neutron Research (NCNR) using the emiT apparatus~\cite{rf:BowlesNBS,rf:LisingPRC,rf:emiTRSI}.  The $D$-coefficient is uniquely sensitive to the relative phase of vector and axial vector amplitudes and is also sensitive to scalar and tensor currents. Our result, first reported in reference~\cite{rf:emiTPRL}, sets a new limit on 
this phase
as well as on certain combinations of scalar and tensor currents. 

This paper presents the details of the experiment and  analysis and examines the implications of this result.
The paper is organized as follows: The context of the measurement is presented in the remainder of this introduction. In section~\ref{sec:emiTExperiment} the measurement principle based on the symmetries of the apparatus is presented. The Monte Carlo simulations used to understand the apparatus and important details of the performance of the apparatus  are provided in sections~\ref{sec:MonteCarlo} and~\ref{sec:Apparatus}. In section~\ref{sec:Data} we present the details of the data set; and section~\ref{sec:Systematics} is a detailed discussion of all systematic effects and corrections applied to the data. In section~\ref{sec:Results} we describe the principle of the blind analysis and present the final result; and in section~\ref{sec:Conclusion} we summarize the implications of our result and prospects for improved measurements.

\subsection{Polarized neutron decay}

Neglecting recoil-order corrections, the general form of the Hamiltonian for  beta-decay ({\it e.g.}~ $n\rightarrow p e\bar\nu$) can be written~\cite{rf:JTW}
\begin{eqnarray}
H_{int}&=&(\bar\psi_p \gamma_\mu \psi_n)(C_V\bar\psi_e \gamma_\mu \psi_\nu+C_V^\prime\bar\psi_e  \gamma_\mu \gamma_5 \psi_\nu)\nonumber\\
&-&(\bar\psi_p \gamma_\mu\gamma_5 \psi_n)(C_A\bar\psi_e \gamma_\mu\gamma_5\psi_\nu+C_A^\prime\bar\psi_e \gamma_\mu \psi_\nu)\nonumber\\
&+&(\bar\psi_p \psi_n)(C_S\bar\psi_e \psi_\nu+C_S^\prime\bar\psi_e \gamma_5  \psi_\nu)\nonumber\\
&+&\frac{1}{2}(\bar\psi_p \sigma_{\lambda\mu} \psi_n)(C_T\bar\psi_e \sigma_{\lambda\mu} \psi_\nu+C_T^\prime\bar\psi_e  \sigma_{\lambda\mu}\gamma_5
\psi_\nu)\nonumber,\\
&+&h.c.
\label{eq:JTWHamiltonian}
\end{eqnarray} 
where the subscripts $V$, $A$, $S$, and $T$ respectively refer to vector, axial-vector, scalar, and tensor contributions, and we have left out the pseudoscalar amplitude, which vanishes in the limit of non-relativistic nucleons.
Allowing the possibility of T violation, the $C$'s are complex numbers, and there are 19 free parameters plus a single arbitrary overall phase. The Standard Model is written with only left-handed $V$ and $A$ interactions with $C_V=C_V^\prime$ and $C_A=C_A^\prime$. Thus the number of free parameters is reduced to three: $|C_V|$, $|C_A|$ and the relative phase of $C_A/C_V=|\lambda|e^{i\phi_{AV}}$. The value of $|C_V|=G_F|V_{ud}|$ follows from the conserved-vector-current hypothesis (CVC) 
with $G_F$ determined from the muon lifetime, and $|V_{ud}|$ most precisely determined from super-allowed beta decays~\cite{rf:Hardy05}. The parameter $|\lambda|$ is determined from other measurements including the beta asymmetry ($A$-term) and electron-neutrino correlation ($a$-term).

The Hamiltonian of Eq.~\ref{eq:JTWHamiltonian} leads to the following differential decay rate for polarized neutrons and no final-state electron polarization~\cite{rf:JTW}:
\begin{equation}
\begin{split}
\frac{dW}{dE_ed\Omega_ed\Omega_\nu}=S(E_e)[1+a\frac{\vec p_e\cdot\vec p_\nu}{E_eE_\nu} + b\frac{m_e}{E_e} \\
+\vec P\cdot(A\frac{\vec p_e}{E_e}+B\frac{\vec p_\nu}{E_\nu}+D\frac{\vec p_e\times \vec p_\nu}{E_eE_\nu}),
\end{split}
\label{eq:JTW}
\end{equation}
where $S(E_e)=F(E_e)p_eE_e(E^{max}- E_e)^2$ is the phase space factor with $F(E_e)$ the Fermi-function for $Z=1$, $\vec p_e (E_e)$ and $\vec p_\nu (E_\nu)$ are the momentum(energy) of the electron and antineutrino, and the neutron polarization $\vec P=\frac{\langle\vec J_n\rangle}{J_n}$ is the ensemble average of the neutron spin. The triple-correlation, $D{\langle\vec J_n\rangle}\cdot({\vec p_e}\times {\vec p_\nu})$ is P-even but odd under motion reversal. Time-reversal is  the combination of motion-reversal and initial/final-state reversal.  Thus contributions to $D$ can originate from T-violating interactions and from  final state effects, {\it i.e.}
$D=D_{ {\rm T}\!\!\!\! \slash}+D_{FSI}$~\cite{rf:Holstein}, where
\begin{equation}
D_{FSI}\approx 1.1\times 10^{-5}\ \frac{p_e}{p_e^{max}} +0.3\times 10^{-5}\ \frac{p_e^{max}}{p_e}.
\end{equation}
  For the neutron $D_{FSI}\approx 1.2\times 10^{-5}$ and can be calculated to 1\% or better~\cite{rf:JTWRadCorr,rf:CallanTriemanFSI,rf:AndoFSI}. 

In the neutron rest frame, ${\vec p_\nu}=-({\vec p_p+\vec p_e})$ (for cold neutrons, the neutron velocity in the lab frame has a negligible effect), and the triple correlation can be written
 \begin{equation}
 D\langle\vec J_n\rangle\cdot({\vec p_e}\times {\vec p_\nu})= D\langle\vec J_n\rangle\cdot({\vec p_p}\times {\vec p_e}).
\end{equation}
We therefore extract $D$ by measuring proton-electron angular correlations in polarized-neutron decay.
 
\subsection{The Physics of $D_{ {\rm T}\!\!\!\! \slash}$} 

In the context of Eq.~\ref{eq:JTWHamiltonian},  $D_{ {\rm T}\!\!\!\! \slash}$ depends on the values of $C_i$ and can be written as~\cite{rf:JTWRadCorr}
\begin{eqnarray}
\xi D_{ {\rm T}\!\!\!\! \slash}=
-\frac{2}{\sqrt{3}}|M_{F}||M_{GT}|
{\rm Im}
[
C_VC_A^{*}+C_V^{\prime} C_A^{\prime *}\nonumber
]\\
+
\frac{2}{\sqrt{3}}|M_{F}||M_{GT}|\nonumber
{\rm Im}
[
C_SC_T^{*}+C_S^{\prime} C_T^{\prime *}
]\nonumber
\\
-\frac{2}{\sqrt{3}}|M_{F}||M_{GT}|
\frac{\alpha m_e}{p_e} {\rm Re}[C_SC_A^{*}+C_S^{\prime}C_A^{\prime *}\nonumber
\\
 -C_VC_T^{*}-C_V^{\prime} C_T^{\prime *}],\nonumber\\
\end{eqnarray}
where $|M_F|$ and $|M_{GT}|$ are matrix elements for Fermi and Gamow-Teller transitions (for neutrons, $|M_{GT}|^2/|M_F|^2=3$), and 
\begin{eqnarray}
\xi&=&|M_F|^2(|C_V|^2+|C_V^\prime|^2+|C_S|^2+|C_S^\prime|^2)\nonumber\\
&+&|M_{GT}|^2(|C_A|^2+|C_A^\prime|^2+|C_T|^2+|C_T^\prime|^2).
\end{eqnarray}
We assume  terms quadratic in $C_S$ and $C_T$ can be neglected in $\xi$, and 
take $C_V=C_V^{\prime}$ and $C_A=C_A^{\prime}$ ({\it i.e.} left-handed $V$-$A$ interactions) so that
\begin{eqnarray}
D_{ {\rm T}\!\!\!\! \slash}\approx \frac{1}{ 1+3|\lambda|^2} \{-2\frac{{\rm Im}(C_VC_A^*)}{ |C_V|^2}+ \frac{{\rm Im}(C_SC_T^*+C_S^\prime C_T^{\prime *})}{|C_V|^2}\nonumber\\
+\frac{\alpha m}{p_e}{\rm Re}(\lambda^*\frac{C_T^*+C_T^{\prime *}}{C_A^*}-\lambda^*\frac{C_S+C_S^\prime}{C_V})\},\nonumber\\
\label{eq:DVAST}
\end{eqnarray}
where $\lambda=|\lambda| e^{i\phi_{AV}}=C_A/C_V$.
The first term  is sensitive to the phase of $\lambda$ and can be approximately written $D^{\rm VA}_{ {\rm T}\!\!\!\! \slash}\approx 0.435\sin\phi_{AV}$ for $|\lambda |=1.2694$~\cite{rf:PDG2010}. The remaining terms show sensitivity to combinations of scalar and tensor amplitudes that signal beyond-Standard-Model Physics.

Standard-Model CP violation makes very small contributions to $D_{{\rm T}\!\!\!\!\slash}$. 
The CKM phase  enters light-quark processes at loop level yielding $D_{ {\rm T}\!\!\!\! \slash}\approx 10^{-12}$~\cite{rf:KhriplovichThetaQCD,rf:Herczeg}. The parameter $\theta_{QCD}$ is tightly constrained by the EDMs of the neutron and $^{199}$Hg and also makes a very small contribution to  $D_{ {\rm T}\!\!\!\! \slash}$, {\it i.e.} less than  $10^{-14}$~\cite{rf:KhriplovichThetaQCD,rf:Herczeg}. Beyond-Standard-Model physics, in particular left-right symmetric models, exotic fermions, and leptoquarks give rise to couplings that contribute to $D_{ {\rm T}\!\!\!\! \slash}$ at a level comparable to the current experimental sensitivity; however, it has been argued that limits on T-odd/P-odd observables, in particular EDMs, can be used to place limits on T-odd/P-even interactions and thus on $D_{\not T}$ by calculating the T-odd/P-even effects of radiative loops on T-odd/P-odd interactions~\cite{rf:Khriplovich91,rf:Conti92,rf:Haxton94,rf:Kurylov01,rf:SeanTulinPC}.  
For the neutron EDM, this limit is more stringent than the sensitivity of the experiment described here by as much as an order of magnitude; however, this argument is  based on the assumption of complete absence of cancellations between different contributions to the neutron EDM, which cannot be {\it a priori} excluded~\cite{rf:SeanTulinPC}.  Table~\ref{tb:Models} summarizes the contributions to $D_{\not T}$ from the Standard Model  and extensions.

\begin{table}[htpb]
\caption{Expected contributions to $D_{ {\rm T}\!\!\!\! \slash}$ for neutron decay from parameters of the Standard Model and beyond-Standard-Model physics based on measurements in other systems. The broad range of limits arises in the cases of significant model dependence. }
\begin{ruledtabular}
\begin{tabular}{lc}
Model/Parameter & Limit on $D_{ {\rm T}\!\!\!\! \slash}$  \\
\hline
\\ [-8 truept]
CKM phase & $10^{-12}$  \\
$\theta_{QCD}$ & $2\times 10^{-15}$\\
Left-right symmetry & $10^{-7}$-10$^{-5}$ \\
Non-SM fermions & $10^{-7}$-10$^{-5}$ \\
Charged Higgs SUSY & $10^{-7}$-10$^{-6}$  \\
Leptoquark &  10$^{-5}$-10$^{-4}$  \\
\end{tabular}
\end{ruledtabular}
\label{tb:Models}
\end{table}

\subsection{Recent results}

The two most recent measurements of $D$ in neutron decay are the  emiT-I measurement and  the TRINE experiment result from Institute Laue Langevin, Grenoble (ILL). For emiT-I,  \mbox{$D=[-6\pm 12({\rm stat})\pm 5({\rm sys})]\times 10^{-4}$~\cite{rf:LisingPRC}}, and the TRINE result was \mbox{$D=[-2.8\pm 6.4 ({\rm stat})\pm 3.0 ({\rm sys})]\times 10^{-4}$~\cite{rf:SoldnerTRINE}}. The current Particle Data Group average for the neutron, $D=[-4\pm 6]\times 10^{-4}$~\cite{rf:PDG2010}, also includes  results from references~\cite{rf:Steinberg74,rf:Erozolim74,rf:Erozolim78}. A measurement in $^{19}$Ne, where the final-state-interactions are more than an order of magnitude larger than for the neutron, 
resulted in $D_{\rm ^{19}Ne}=[0.7\pm 6]\times 10^{-4}$~\cite{rf:19Ne1,rf:19Ne2}. We also note that the $R$ coefficient of the T-odd/P-odd correlation $\vec J_n\cdot({\vec p_e}\times \vec\sigma_e)$, which is linearly sensitive to S and T amplitudes, has recently been measured for the neutron~\cite{rf:Kozela} and for $^8$Li~\cite{rf:Stromecki}.

\section{The emiT-II experiment}
\label{sec:emiTExperiment}
 
\begin{figure*}
\begin{centering}
\centerline{\includegraphics[width=6.5 truein]{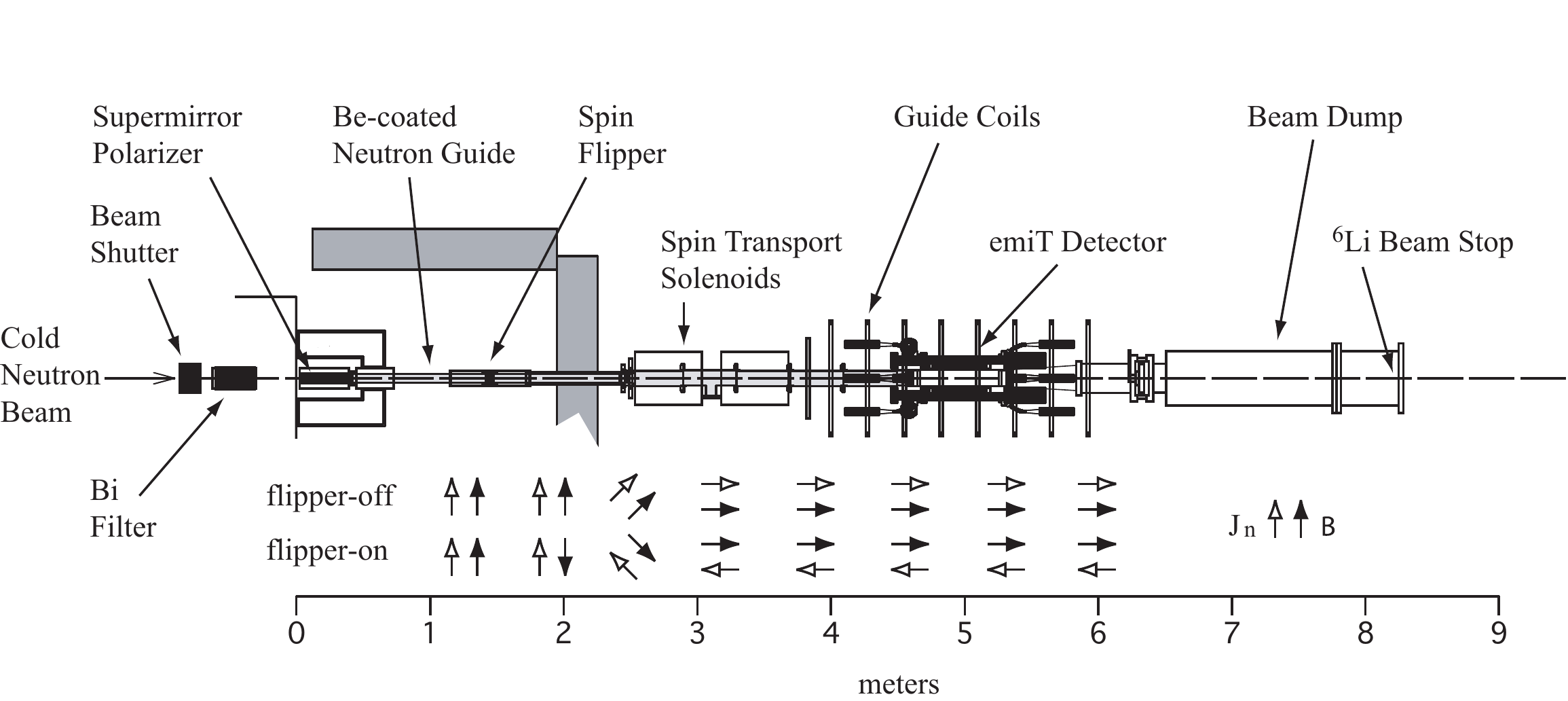}} 
\caption{Layout of the emiT-II experiment beam-line. The neutron beam is nominally unpolarized upstream of the polarizer, and is vertically polarized downstream. Polarized neutrons are guided within beryllium coated glass tube to the detector.   
As shown, the spin flipper reverses the direction of the vertical magnetic field $\vec B$ over a short distance so that the neutron spin $\vec J_n$, which remains polarized vertically upward, reverses with respect to the magnetic field. Downstream of the spin flipper, solenoids rotate the magnetic field  into the horizontal direction, parallel to the neutron beam.  
\label{fg:emiTLayout}}
 \end{centering}
 \end{figure*}

The emiT experiment was designed to measure proton-electron coincidences in the decay of neutrons polarized along the axis of an array of detectors. The symmetry of the detector array allowed us to discriminate the triple correlation from the T-even/P-odd $A$- and $B$-coefficient correlations.  
The layout of the experiment is shown in Fig.~\ref{fg:emiTLayout}. The cold neutron beam was transported by the neutron guide NG6 to the experiment. The neutrons were polarized  and passed through a  spin flipper. Downstream of the spin-flipper, neutron spins were adiabatically transported through rotation of the magnetic field to longitudinal, along the axis of the detector array.

\begin{figure*}
\begin{centering}
\vskip -1.5 truein
\centerline{\includegraphics[width=7.9 truein]{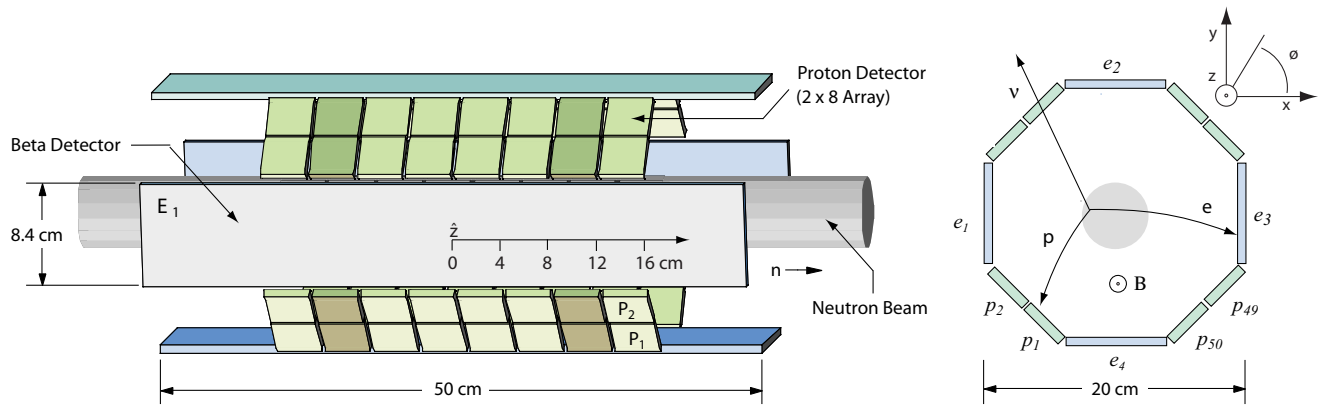}}
\vskip -2. truein
%\centerline{\includegraphics[width=5 truein]{Detector3D_V3_side}\hskip -0.25 truein\includegraphics[width=2 truein]{Detector3D_V6_end}}
\caption{The emiT-detector array. Left: side view showing proton-detector planes, with 16 proton cells (2$\times$8) in each plane, and 50 cm long electron detectors. Right: end view showing the four proton-detector planes and four electron detectors.  The magnetic field,  directed parallel to the average neutron velocity, causes the proton and electron trajectories to be curved as indicated by the greatly exaggerated paths shown. In this paper, we refer to the magnitude of the relative angles between a proton detector and electron detector, {\it e.g.}~ for proton detector $p_1$, these are  35$^\circ$, 55$^\circ$, 125$^\circ$, and 145$^\circ$,  for
 $e_4$, $e_1$, $e_3$, and $e_2$ respectively. For reference, distances are measured from the center of the detector array.
\label{fg:emiTDetector}}
 \end{centering}
 \end{figure*}

\begin{figure}
\vskip -.5 truein\centerline{\includegraphics[width=4 truein]{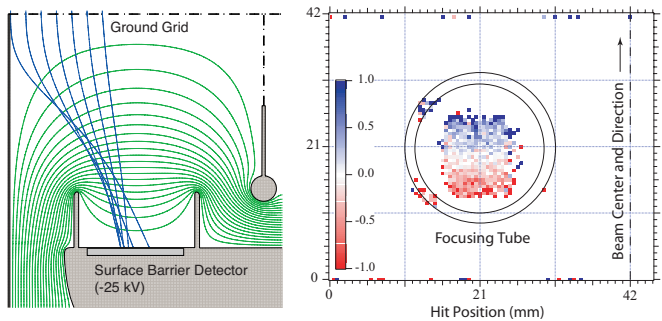}}\vskip -.7 truein
%\centerline{\includegraphics[width=1.4 truein]{FocusingCell}\hskip 0.1 truein\includegraphics[width=1.7 truein]{HitDiff2}}
\caption{Left: proton-cell layout with calculated proton paths showing focussing properties of electrode geometry. Right: Monte Carlo generated hit pattern for neutron spin parallel (red) and opposite (blue) $\hat z$.\label{fg:ProtonCell}}
\end{figure}

The detector array, illustrated in Fig.~\ref{fg:emiTDetector}, consists of four electron-detectors alternating with four proton-detector planes arranged in an octagonal geometry concentric with the neutron beam. Each of the four proton-detector planes consists of 16 separate cells arranged in two
 rows of eight cells. The protons are detected by negatively-biased surface barrier detectors (SBDs) that are incorporated into a focusing cell illustrated in Fig.~\ref{fg:ProtonCell}. Within the fiducial volume of the detector array, neutrons are polarized parallel or antiparallel to the magnetic field depending on the state of the  spin flipper. The magnetic field in the detector region is approximately 560~$\mu$T in magnitude and is 
  aligned parallel to the neutron beam and detector axis.

\subsection{Electron-proton Coincidence Events}

The data set consists of 512 sets of coincidence events from the combination of 64 proton cells and four electron detectors for the two spin-flipper states. 
The total number of counts for a given run time is labeled as $N_{\pm}^{p_i e_j}$, where the $\pm$ indicates the spin-flipper state (neutrons nominally parallel/opposite to the magnetic field $\vec B$),
$p_i$ labels the proton-cell, and $e_j$ labels the electron detector. 
The neutron-spin dependence of the count rates  depends on the correlations given by $A$, $B$, and $D$ and is given by the difference of rates for the two spin-flipper states, while the  average of rates includes the spin-independent beta-neutrino correlation ($a$ term). 

In order to isolate the neutron-spin dependent terms, we define the asymmetry 
\begin{equation}
w^{p_i e_j} = \frac{N^{p_i e_j}_{+}-N^{p_i e_j}_{-}}{N^{p_i e_j}_{+}+N^{p_i e_j}_{-}}. 
\end{equation}
In principle, the $N_{\pm}^{p_i e_j}$ follow from integrating Eq.~\ref{eq:JTW} over the neutron beam, the detectors' acceptances, electron momentum, and neutrino angles for a fixed time so that
\begin{eqnarray}
w^{p_i e_j}\! \approx\!
\frac{
A\langle \beta_e \vec P\!\cdot\! \hat p_e\rangle\! +\! B\langle  \vec P\!\cdot\! \hat p_\nu\rangle\!+\! D\langle \beta_e(\frac{p_p}{p_\nu}) \vec P\!\cdot\! (\hat p_p\times \hat p_e)\rangle 
}{
\langle 1\rangle +  a \langle \beta_e {\hat p_e\cdot\hat p_\nu}\rangle} \nonumber,\\
\label{eq:ws}
\end{eqnarray}
where $\beta_e=v_e/c$ is the electron velocity, $\vec P$ is the neutron polarization at a given position, and the  brackets ($\langle\ \rangle$) indicate that each term is integrated over energies, the neutron-beam distribution, and solid angles for proton detector $p_i$ and electron detector $e_j$. As shown below, the $D$-coefficient term can be isolated by forming a specific combination of $w^{p_ie_j}$  that cancels the parity-violating $A$ and $B$ correlations.

\subsection{The ideal experiment}
\label{sec:IdealExperiment}

In order to understand the analysis technique, we begin by considering an ideal experiment with uniform longitudinal neutron polarization ($\vec P=P\hat z$), uniform neutron-beam density, and uniform efficiencies for all proton and electron detectors. Consider a proton detected in $p_1$ in coincidence with an electron detected in either  detector $e_2$ or $e_3$ as shown in Fig.~\ref{fg:emiTDetector}. (Coincidence rates of $p_1$ with $e_1$ and $e_4$ were about 15-25 times lower and were not used to extract $D$ in this analysis.) Thus, for longitudinal neutron polarization, the asymmetry from Eq.~\ref{eq:ws} can be written
\begin{eqnarray}
w^{p_ie_j} \approx P\kappa^{p_ie_j}[D\langle \beta_e(\frac{p_p}{p_\nu}) \hat z\cdot(\hat p_p\times \hat p_e)\rangle\nonumber\\
+ A \langle\beta_e\cos\theta_e\rangle+B \langle\cos\theta_\nu\rangle],
\label{eq:w_L}
\end{eqnarray}
where $\theta_e$ and $\theta_\nu$ are the polar angles of the electron and antineutrino  with respect to the neutron polarization $\vec P$, and
\begin{equation}
\nonumber
\kappa^{p_ie_j}=\frac{1}{\langle 1\rangle + a\langle \beta_e {\hat p_e\cdot\hat p_\nu}\rangle}.
\end{equation}
It is useful to define an instrumental constant that characterizes the sensitivity of $w^{p_ie_j}$ to the triple correlation, {\it i.e.} $K_D^{p_ie_j}=\partial w^{p_ie_j}/\partial D$ or
\begin{equation}
 K_D^{p_ie_j}=\kappa^{p_ie_j}\langle \beta_e(\frac{p_p}{p_\nu}) \hat z\cdot(\hat p_p\times \hat p_e)\rangle.
\end{equation}
The $K_D^{p_ie_j}$ used in the analysis were determined by Monte Carlo studies and are discussed in section~\ref{sec:KDerror}.

In order to isolate the triple correlation, we note that  the longitudinal component of the cross product, $\hat z\cdot(\hat p_p\times \hat p_e)$,
has opposite sign for $p_1e_3$ and $p_1e_2$ coincidences: for $p_1e_3$, it is positive, 
 and for $p_1e_2$, it is negative. Thus 
 we  form a difference of spin-flip asymmetries for the two electron detectors. For example, for $p_1$ we have: \mbox{$v^{p_1}=\frac{1}{2}(w^{p_1e_3}-w^{p_1e_2})$. } Using Eq.~\ref{eq:w_L}, the $v^{p_i}$ for longitudinal polarization are
  \begin{eqnarray}
v^{p_1}
&\approx& \bar K_DPD\nonumber\\
&+&P\frac{A}{2}[ \kappa^{p_1e_3}\langle\beta_e\cos\theta_e\rangle^{p_1e_3}-\kappa^{p_1e_2}\langle\beta_e\cos\theta_e\rangle^{p_1e_2}]\nonumber\\
&+&P\frac{B}{2} [\kappa^{p_1e_3}\langle\cos\theta_\nu\rangle^{p_1e_3}-\kappa^{p_1e_2}\langle\cos\theta_\nu\rangle^{p_1e_2}],
\label{eq:v_L}
\end{eqnarray}
where $\bar K_D=\frac{1}{2}(K_D^{p_1e_3}-K_D^{p_1e_2})$. 
Due to strong anti-correlation of proton and electron momenta,  the asymmetries depend strongly on the axial position of the proton cell. (Data are shown in Figs.~\ref{fg:ws} and~\ref{fg:vs}.)  For example, assuming $D=0$, the $v^{p_{j}}$
(and $w^{p_{i}e_{j}}$)
 are equal but opposite for the upstream proton cell and an axially symmetric downstream proton cell for a uniform neutron beam, {\it e.g.}~ $v^{p_1}=-v^{p_{15}}$. Also note that in the absence of the 560~$\mu$T magnetic field, the $v^{p_i}$ are opposite for the adjacent proton cell, {\it e.g.}~ $v^{p_1}=-v^{p_2}$. Thus, for an ideal experiment with uniform longitudinal polarization and beam, the 
average of $v^{p_i}$ from an upstream-downstream  pair of proton cells  ({\it e.g.}~  $v^{p_1}$ and $v^{p_{15}}$)  or adjacent cells ({\it e.g.}~ $v^{p_1}$ and $v^{p_2}$) will cancel the beta-asymmetry and neutrino-asymmetry terms leaving the $D$-coefficient term; however, the magnetic field affects the average $\beta_e$ differently for odd ({\it e.g.} $p_1$) and even ({\it e.g. $p_2$}) proton detectors. Thus we combine data for all four proton cells in order to isolate $D$.

\subsubsection*{Transverse neutron polarization}

A small misalignment of the magnetic field with respect to the detector axis gives rise to transverse-polarization effects. In the limit of a small beam diameter and a small angular acceptance of the proton cell, the average proton momentum $\langle\vec p_p\rangle$ is the same for the two electron detectors ({\it e.g.}~$e_3$ and $e_2$), and for transverse polarization, indicated by the subscript $T$,  we can write 

\begin{equation}
\begin{split}
v_T^{p_i} \approx 
 \ \  &v_T\sin(\phi_P-\phi^{p_i}),
\end{split}
\label{eq:v_T}
\end{equation}
where  $\phi_P$ is the azimuthal angle of the polarization, $\phi^{p_i}$ is the effective azimuthal position of the proton cell as indicated in Fig.~\ref{fg:emiTDetector}, 
and $v_T\approx 0.46 P\sin\theta_P$.  The factor 0.46 is consistent with both Monte-Carlo simulations and the  transverse-polarization calibration runs discussed in section~\ref{sec:ATP}.  Thus, for uniform neutron density and polarization, the sinusoidal dependence of $v_T^{p_i}$ on the azimuthal position of the proton-cell averages to zero when data from all four proton-detector planes are combined.

\subsubsection*{Extracting $D$}

In order to cancel the effects of transverse polarization, the  beta and neutrino asymmetries,  and the 560~$\mu$T magnetic field, we must combine the $v^{p_i}$ from at least 16 proton cells symmetrically located with respect to the center of the detector: two adjacent cells upstream of the detector center paired with two adjacent  downstream cells from all four proton planes. Each set of 16 proton cells have the same  $|z^{p_i}|$, {\it i.e.} $z^{p_i}$ = $\pm$2 , $\pm$6, $\pm$10, and $\pm$14 cm. For example, the shaded detectors in Fig.~\ref{fg:emiTDetector} correspond to $z^{p_i}=\pm 10$ cm. For each $|z^{p_i}|$ we define a measured quantity $\tilde D$ given by
\begin{equation}
\tilde D= \frac{1}{\bar K_D P}\sum_{|z^{p_i}|={\rm const}}v^{p_i},
\label{eq:Dtilde}
\end{equation}
where $\bar K_D=0.378\pm 0.019$ is discussed in section~\ref{sec:KDerror}. Due to the symmetry with respect to the detector center, each of the four possible $\tilde D$  is an independent measurement of the same nominal quantity. This provides cross checks and maximum statistical power. The $\tilde D$ are subject to a variety of systematic effects that are estimated and applied as corrections in order to determine $D$. Corrections to $\tilde D$ are discussed in section~\ref{sec:Systematics}.

\section{Monte Carlo simulations}
\label{sec:MonteCarlo}

Monte Carlo simulations were used both in the design of the experiment
and  to estimate  several of the systematic effects that impact the analysis. Separate simulations addressed the detector responses, proton focusing cells, and neutron spin transport.

To track electrons, we used  {\scshape penelope}~\cite{rf:penelope}, which has been experimentally validated for a
variety of kinematics of relevance to neutron decay~\cite{rf:Martin2006}. 
To track protons we used custom code embedded into {\scshape penelope}. 
The Monte Carlo code used  the measured beam distributions and  
took into account the 560~$\mu$T axial magnetic field and details of the 
detector as described below. 

Simulations used a detector model that included all surfaces visible to decaying particles including cryo-panels and the
proton-cell ground planes. For simplicity in backscattering studies,
the grids on the proton detectors were replaced in the model with solid foil, and the simulations were verified for this substitution.
The {\scshape penelope}-based Monte Carlo code was used to analyze the effects of the electron response function and
backscattering of protons and electrons.
Section~\ref{sec:Backscattering} provides details on how the
backscattering events were identified. Section~\ref{sec:Nonuniformbeta} gives
details on the issues associated with the scintillator response.

A separate simulation using  {\scshape simion}~\cite{rf:SIMION}  was developed to track proton trajectories within the proton cells for proton events generated by the {\scshape penelope}-based Monte Carlo. The detailed proton-cell geometry  was used to determine the electric fields and to analyze focusing efficiency as a function of incident proton momentum and position on the focusing-cell grid. In order to minimize computer time, protons that penetrated beyond the grids were assumed to be detected with efficiencies calculated 
with these separate  proton-cell simulations. Section~\ref{sec:Backscattering} describes issues related to proton backscattering.

The energy-loss code {\scshape SRIM}~\cite{rf:SRIM} was used to provide proton energy loss backscattering probabilities.
In addition, SBD models were used with {\scshape SRIM} calculations to evaluate the response, energy loss, and build up of condensation on the SBDs.

The propagation of neutron spin components from the polarizer and through the spin flipper and guide field were modeled by numerical integration of the Bloch equations using Monte Carlo techniques to generate neutron trajectories~\cite{rf:SRHThesis}. 

\section{Apparatus peformance}

\label{sec:Apparatus}

Detailed descriptions of the experiment components, including beam line, polarization, spin flipper, spin-transport magnetic fields and the detector array for the emiT-II run are described in detail in references~\cite{rf:LisingPRC,rf:emiTRSI,rf:LisingThesis,rf:SRHThesis,rf:HPMThesis}. Here we summarize specific features that impact the analysis and systematic effects described in this paper.

\subsection{Beam}

\label{sec:BeamDistribution}

Lead-backed lithium-glass collimators defined the beam and reduced backgrounds at the detector.  The beam dump was located two meters downstream of the detector and was also composed of lithium glass. A 3 mm hole in the glass allowed neutrons to pass through and the beam rate to be monitored continuously throughout the run. The mean neutron capture flux (the average of the neutron flux weighted by the inverse neutron velocity, $1/v_n$) was monitored with a fission monitor~\cite{rf:FissionMonitor}.

\begin{figure}
\begin{centering}
\centerline{\includegraphics[width=3.25 truein]{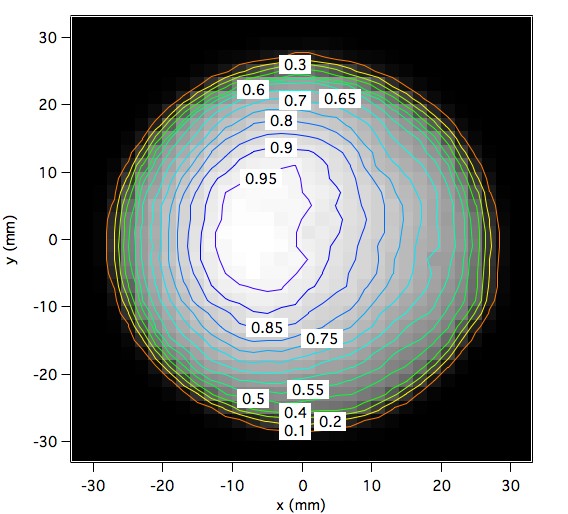}}
\centerline{\includegraphics[width=3.25 truein]{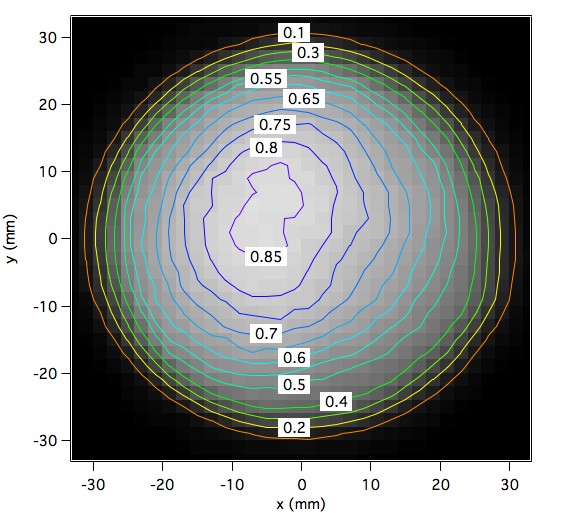}}
\caption{Neutron-beam distributions for positions 18 cm upstream of the center of the detector (top) and 18 cm downstream of the detector center (bottom).   Contours show the  film exposure relative to the maximum at the center of the upstream map.
\label{fg:BeamMaps}}
 \end{centering}
 \end{figure}

Estimates of the distribution of neutron density were used in the Monte-Carlo modeling of the experimental and systematic effects. The cold neutron beam from NG6 was mapped by activating a 25 $\mu$m thick, natural-abundance dysprosium foil. Neutron absorption on $^{164}$Dy (28.2\% abundance) follows the $1/v_n$  law, and the foil is thin enough to provide a sufficiently accurate measure of the neutron density.  The activated foil was subsequently laid on a beta-sensitive film, and the exposure  was measured with an image reader with pixel-resolution of  200~$\mu$m by 200~$\mu$m~\cite{rf:ChengFilmReader}. 
Maps 18 cm upstream and 18 cm downstream of the center of the detector array are shown in Fig.~\ref{fg:BeamMaps}. The dominant features of the maps are the expansion of the beam (by about 5 mm in radius over the 80 cm length of the detector array) and the horizontal shift of the beam, which is due to the properties of the polarizer. The beam expansion affects the longitudinal cancelations of the $v^{p_i}$ from upstream-downstream combinations and leads to a systematic effect discussed in section~\ref{sec:ExpandingBeamCorrection}. 

\subsection{Polarization and spin flipper}
\label{sec:Polarization}

The neutron polarizer is a supermirror bender (PSM)~\cite{rf:PSM}.
Neutrons  are polarized parallel to the vertical magnetic field of the PSM. 
The spin flipper consists of two closely-spaced current sheets with horizontal currents. The magnetic field from the upstream current sheet was parallel to the PSM field, and the current in the downstream current sheet could be set  parallel or antiparallel to the upstream current sheet. For antiparallel current,  the vertical magnetic field reverses direction over a distance of about 1 mm corresponding to a time of about 2 $\mu$s in the rest frame of a cold neutron. This time is short compared to the inverse Larmor frequency, and the neutron spin does not follow the magnetic field adiabatically. The result is that the neutron spin remains oriented in the original upward direction while the magnetic field reverses from up to down, thus reversing the projection of the spin with respect to the magnetic field direction. Small transverse neutron-spin components arise due to imperfect transitions. These transverse components precess around the magnetic guide-field as the neutrons move through the apparatus. Thus the polarization direction of a decaying neutron depends on the vertex of the decay as well as the neutron's velocity and the magnetic field. The distribution of neutron velocities leads to azimuthal averaging of the transverse polarization components. The 560~$\mu$T field was chosen so that the orientation of the neutron polarization $\vec P$ varied by less than $2\times 10^{-3}$ across the detector. Downstream of the spin flipper, the guide field adiabatically rotates the neutron spin into the longitudinal direction and remains longitudinal throughout the detector array. 

\begin{figure}
\begin{centering}
\centerline{\includegraphics[width=3.25 truein]{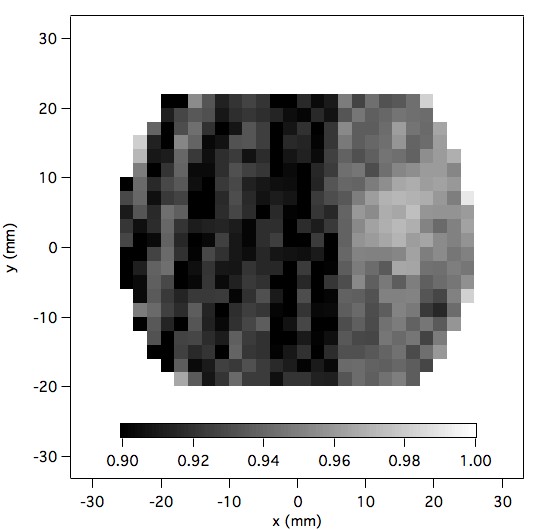}}
\caption{Neutron polarization map showing the position dependence of the estimated lower-limit on neutron polarization assuming $P=A_P$.  The top and bottom are cut off by the collimating mask that defined the aperture of the analyzer.
The increase of polarization from left to right is due to the PSM bending properties.    
\label{fg:PolMap}}
 \end{centering}
 \end{figure}

The polarization was measured and mapped using a second polarizer as an analyzer downstream from the spin flipper~\cite{rf:emiTRSI}. A map of the estimated neutron polarization, assuming a perfect spin flipper and identical polarization and analyzing power ($P=A_P$),  is shown in Fig.~\ref{fg:PolMap}. Though neither of these assumptions is accurate, this provides a lower limit on the neutron polarization~\cite{rf:emiTRSI}. The increase in polarization from left to right is a property of the polarizer and leads to effects that are discussed in section~\ref{sec:Polnonuniformity}. When averaged across the beam, the  lower limit on the neutron polarization is $P>0.91$ (90\% c.l.), and we use $P=0.95\pm 0.05$ in the analysis of $D$. The polarization and spin-flipper characteristics appeared to be quite stable over long periods,  and the measured polarization was consistent with that measured for the first emiT run, which set the  lower limit of $P>93\%$~\cite{rf:LisingPRC}.

\subsection{Magnetic fields}

 \label{sec:Bfieldmap}
 
Within the detector region, the magnetic field is maintained by a set of eight 0.95 m diameter coils uniformly spaced over 2 m. The outermost coils had independently adjustable currents, while the inner six coils were connected in series. The current, alignment, and position of each coil were set to optimize the longitudinal field uniformity. An array of longitudinal coils was deployed as a cosine magnet to cancel uniform transverse magnetic field components. These coils were also used to rotate the magnetic field for the transverse-polarization calibration runs. Additional trimming of non-uniform components of the magnetic field was necessary due to the magnetized steel in the floor and magnetic fields from other instruments in the NCNR guide hall~\cite{rf:SRHThesis,rf:HPMThesis}. The currents in all coils were nominally constant with no active compensation; however, all currents were monitored, and the fields in the vicinity of the decay region were continuously measured using two 3-axis flux-gate magnetometers.  
 Magnet-current and detector-field monitor data were used to define cuts on the data as described section~\ref{sec:Cuts}.

 The magnetic field in the detector region was mapped with a 3-axis fluxgate magnetometer before and after the emiT-II run. The large axial field of 560~$\mu$T made precise measurement  of transverse component difficult due to flexing of the support rail for the magnetometer; however, it was possible to estimate transverse components of 1-3 $\mu$T. The corresponding magnetic-field misalignment was measured to be as large as 5.4 mrad, which was consistent with estimates based on the transverse-polarization calibration runs described in section~\ref{sec:ATP}.

\subsection{Detector overview}

The detector array is shown schematically in Fig.~\ref{fg:emiTDetector}. The main detector chamber was milled from a single block of aluminum.  The chamber was supported on kinematic mounts and included mounts for cross-hair assemblies at both the upstream and downstream ends. Beam-line and detector components were mechanically positioned and aligned to better than 1 mm and 2 mrad, and the alignment was checked with crosshairs during final assembly.   

The vacuum during data taking was maintained using a cryopump just downstream of the main detector chamber. In addition, liquid-nitrogen cooled cryopanels were situated at the ends of the proton detector assembly.  The vacuum measured in this region was typically in the range of 3-4$\times10^{-7}$ Torr. This low pressure significantly reduced backgrounds due to  neutron interactions with residual gas and minimizes the possibility of scattering and neutralization of decay protons.   On the other hand, changes in proton detector dead-layer of roughly 3\% per month, most likely due to  the build-up of water and other volatile materials, were observed.

\subsection{Proton detectors}
\label{sec:protondetectors}
The proton-detector cell is illustrated in Fig.~\ref{fg:ProtonCell}. Each cell consisted of a grounded box with the top and the upper half of the sides covered by a grounded wire mesh (97\% transmitting) through which the recoil protons enter. Once inside the box, the protons were accelerated and focused onto the 
SBD (Ortec model AB-020-300-300-S) by the field produced by a cylindrical tube maintained at a negative potential with respect to ground. During the course of the experiment, the acceleration voltage  was varied in the range $25$\,kV to $31$\,kV.  Geometric constraints required that the SBD be positioned off-center relative to the focusing tube.  In addition, the cryogenic epoxy holding the silicon wafer into its mount varied in thickness by up to one millimeter around the edge of the detector ring. 
These and other imperfections  were included in the Monte-Carlo model of the proton focusing discussed in Section~\ref{sec:MonteCarlo}. The proton SBD active layer was 300 $\mu$m thick and 300\,mm$^2$ in area.

The proton detectors were periodically calibrated {\it in situ} with $^{241}$Am and $^{109}$Cd gamma-ray sources. Typical detector resolution was 4 keV FWHM but varied by several keV from channel to channel and over time.  In most channels,  thresholds were adjusted to reduce the count rates. This also resulted in truncating part of the accelerated proton spectrum introducing a proton-energy dependent efficiency and systematic effect discussed in section~\ref{sec:CoulterEffect}. Low energy tails on the proton peaks, observed in some channels, were of particular concern because of the thresholds.  
The source of these tails was inconclusive and may have been due to a combination of a number of possibilities including dead-layer buildup and scattering from the focusing tubes themselves. 
 In addition, reversible build-up of the detector dead layers was observed to occur over time. 
 
 During the experiment detectors showed a variety of problems, including  high leakage current and breakdown, and a portion of the data were taken with fewer than 64 operating SBDs.
Over the course of the experiment, however, only a few SBDs were not operational at any one time, and when averaged over the entire run, every proton cell had a duty factor greater than about 90\%.

\subsection{Electron detectors}

Electron detectors are 50 cm by 8.4 cm by 0.64 cm thick BC408 plastic scintillators with Burle-8850 photomultiplier tubes (PMTs) at each end. 
The sensitive region of the electron detectors was fabricated from slabs of plastic scintillator cast to a thickness of 0.64\,cm and diamond milled to a rectangular prism measuring 50\,cm by 8.4\,cm. The thickness was sufficient to stop a 1-MeV electron and was therefore adequate for detecting neutron decay electrons, which have an endpoint energy of 782\,keV. The  scintillators were wrapped in aluminized mylar and 20 $\mu$m thick aluminum foil to prevent charging and to shield the detectors from x-rays and field-emission electrons. This thickness of aluminum stopped electrons of energy up to about 50 keV at normal incidence and led to energy loss of 20 keV for incident electrons with energy 200 keV. The foil and mylar were included in all Monte Carlo simulations.

 Scintillation photons guided to either end of the scintillator by total internal reflection at the smooth surfaces were detected by the PMTs.  An electron event requires coincidence of both PMTs within a timing window of 100 ns. The time difference of the phototube signals at the two ends of each scintillator was available to determine the electron position with resolution of about 10 cm but was not used in the analysis.

The electron detectors were calibrated periodically during the run using a $^{207}$Bi source. Conversion electrons from the source passed though a thin Kapton window into the vacuum chamber and were incident on the backside of the scintillator slabs passing through a hole in the mylar and aluminum foils.    Gain drifts of up to 3\% per month were observed, and PMT base voltages were occasionally adjusted to stabilize the gains.  Remaining drifts were compensated run-by-run in the analysis using features of the neutron-decay-electron-energy spectrum. A threshold of 80\,keV detected electron energy was applied to the data in order to eliminate the effects of residual gain drifts. 
The effects of remaining small nonuniformities in electron-energy threshold and response are discussed in Section~\ref{sec:Nonuniformbeta}.

\subsection{Monitors}
\label{sec:DAQMonitors}

The following parameters were continuously logged during data taking: 

\begin{enumerate}

\item Magnetic fields at two positions within the detector magnetic-field coils  using 3-axis fluxgate magnetometers

\item Currents in all magnetic field coils including neutron-guide solenoids and longitudinal and transverse detector coils

\item The neutron capture-flux monitor

\item Proton-cell acceleration voltage
 
\item SBD bias voltage and current.

\end{enumerate}

The cuts based on these monitor data were used to test for systematic effects and for cross checks of the results as discussed in section~\ref{sec:CrossChecks}.

\section{Data}
\label{sec:Data}

The neutron-decay data consist of proton-electron coincidence events that record which detectors were hit, amplitude of the pulse in each detector, spin-flipper state, and proton time-of-flight ($t_{ep}$).  Data were acquired from October 2002 to November 2003 with some breaks for
reactor and detector maintenance. The data were separated into runs of up to about four hours duration, and a total of 934 runs are included in the final data set. 

\begin{figure}
\vskip 0.1 truein
\centerline{\includegraphics[width=3.5in]{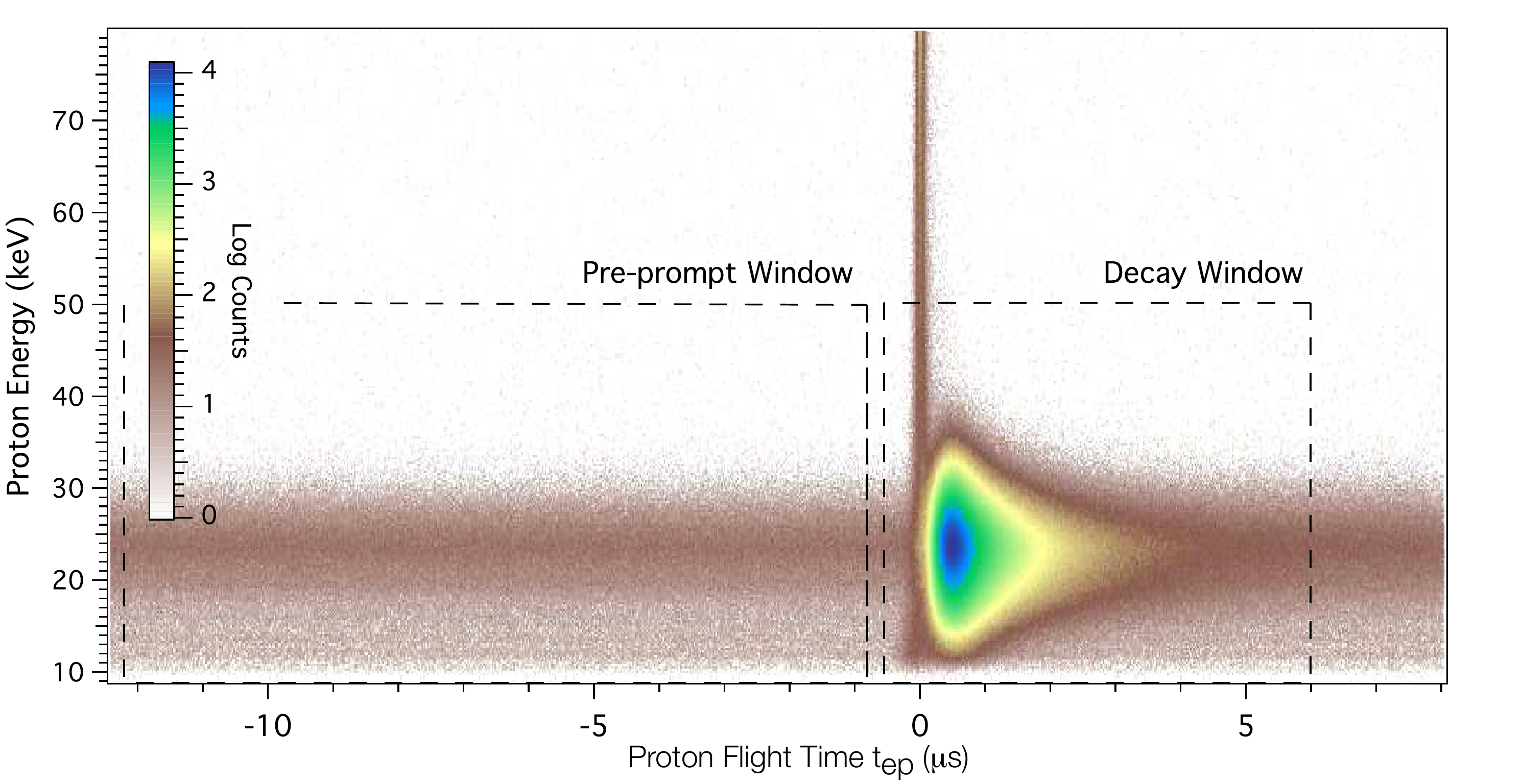}}
\caption{Two-dimensional spectra showing proton energy vs. proton time-of-flight vs. log of counts for all data. The boxes show the cuts used in the analysis (decay window) and to determine the random-coincidence background contribution (pre-prompt window). The pre-prompt events consist primarily of accelerated decay protons in coincidence with an uncorrelated electron-detector hit.
}
\label{fg:ProtonADCvsTOF}
\end{figure}

Fig.~\ref{fg:ProtonADCvsTOF} shows the detected proton energy versus proton time-of-flight for all events. Events prior to $t_{ep}=0$ are random coincidences and are used to estimate background rates as described in section~\ref{sec:Backgrounds}.  The feature near $t_{ep}=0$ primarily consists of true coincidences that are very closely spaced in time compared to neutron-decay-proton-electron events. These prompt coincidences arise from several sources including cosmic-ray muons and neutron capture , which produces  gamma rays that produce a false proton-electron
coincidence through  Compton scattering or pair production. 
In Fig.~\ref{fg:PromptSpectrum} we show the proton-ADC spectrum for the timing window $-0.75\ \mu{\rm s}<t_{ep}<0.12\ \mu{\rm s}$  as a function of detected SBD amplitude.  Protons, primarily from neutron decay are accelerated in the focusing cells and produce the narrow peak at $\approx$ 25 keV. The broad feature at higher energy is the minimum-ionizing peak for relativistic charged particles, {\it e.g.}~from high-energy gamma interactions and cosmic rays. The pre-prompt data were used to estimate the random-coincidence contribution to the backgrounds discussed in section~\ref{sec:Backgrounds}.

Typical count rates were 3 s$^{-1}$ and 100 s$^{-1}$ for single proton and electron detectors, respectively, while the coincidence rate for the entire array was	typically	25 s$^{-1}$. 	A total of $4.7\times 10^{8}$ raw events were acquired. Of these, limits on magnetic field, leakage current, and other monitors removed 12\% of the events, the analysis threshold on electron energy removed 14\% and the single electron hit requirement removed 7\% of the raw events. 
 The data were filtered by cuts on the parameters described in section~\ref{sec:DAQMonitors} as well as on the detected electron-energy and the requirement of a single-electron-detector hit. We also rejected events during a spin flip and during an unpaired spin-flip cycles at the end of a run in order to ensure equal time in each spin-flipper state. 
 
 Most of the data were taken with nominal proton-acceleration voltages of 28 kV with smaller data sets at 25 kV, 27 kV, and 31 kV. The 27-kV data were acquired during the initial running stages in 2002. Discussion of the data and results for different proton-acceleration voltages, running conditions, and detector configurations are discussed in section~\ref{sec:DataSubsetStudies}.

\begin{table}
\caption{Summary of the final event selection.}
\begin{ruledtabular}
\begin{tabular}{lc}
Description & Approximate number\\
\hline
\\ [-8 truept]
Raw events & $4.7\times 10^8$ \\
Monitor cuts & $5.6\times 10^7$ (12\%)\\
Electron-energy threshold &  $6.6\times 10^7$ (14\%) \\
Electron-detector multiplicity &  $3.3\times 10^7$ (7\%)\\
Final event sample &  $3.2\times 10^8$ \\
Random coincidences &  $1.0\times 10^7$ \\
\end{tabular}
\end{ruledtabular}
\label{tb:DataSummary}
\end{table}

\section{Systematic Effects}
\label{sec:Systematics}
 Eq.~\ref{eq:Dtilde} is based on an ideal symmetric experiment, which incorporates the following assumptions:
 a) accurate determination of the number of coincidences for each proton-cell-electron-detector combination, {\it i.e.} accurate background corrections; b)  absence of proton and electron backscattering; c) symmetry of the detector, specifically  the equivalence of proton cells
 and uniform proton and electron detection efficiencies;
d) uniformity of the neutron beam;
e) uniform longitudinal polarization; and
f) accurate determination of $\bar K_D$ and $P$.
A number of effects break the symmetries of the experiment and are accounted for by corrections to $\tilde D$ discussed in detail this section.

\subsection{Backgrounds}
\label{sec:Backgrounds}

A valid event is defined as the coincidence of a proton-detector signal and electron-detector signal that meet the selection criteria on proton and electron energy, proton time-of-flight, and the monitor-data. Electron detection requires a coincidence of both phototubes in one scintillator panel with detected electron energy above 80~keV.  The proton-energy-time-of-flight window is shown in Fig.~\ref{fg:ProtonADCvsTOF}.  Background events that satisfy these criteria arise from several sources. Estimates of the background fractions $\epsilon^{p_ie_j}$  for each component are discussed below, and the averages over all detectors, $\bar\epsilon_b$, are given in table~\ref{tb:Backgrounds}.

 \subsubsection*{Accidental Coincidences}
 
The count rates for each spin-flipper state were corrected  for accidental coincidences on a run-by-run basis by scaling the counts in the pre-prompt timing window ($-12.3\ \mu{\rm s}<t_{ep}<-0.75\ \mu{\rm s}$) shown in Fig.~\ref{fg:ProtonADCvsTOF}. These random coincidences had the expected exponential dependence with respect to proton time-of-flight.  A fit revealed a time constant of $(2690\pm 730)$ $\mu$s, which is sufficiently long that  we made no adjustment for proton time-of-flight. This results in a correction on the background subtraction estimated to be  $(2\pm 2)\times 10^{-3}$.

 \subsubsection*{Prompt Coincidences}
 
\begin{figure}
\centerline{\includegraphics[width=3.5in]{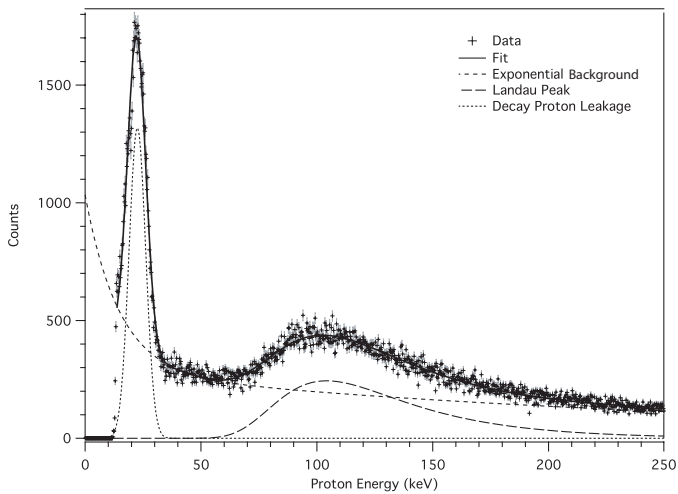}}
\caption{Typical proton-ADC spectrum for the ``prompt'' timing window $-0.75\ \mu{\rm s}<t_{ep}<0.12\ \mu{\rm s}$. The lines show best-fit contributions due to neutron-decay protons, minimum ionizing, and an exponential tail determined from runs with zero proton-acceleration voltage.}
\label{fg:PromptSpectrum}
\end{figure}

An energy spectrum of events in the prompt window (-0.75 $\mu$s to 0.12 $\mu$s) is shown in  Fig.~\ref{fg:PromptSpectrum}. The dominant components are the peak at $\approx$ 25 keV due to the accelerated  protons from neutron decay, minimum ionizing charged particles predominantly due to cosmic-ray muons, and a background continuum. The low-energy behavior of the continuum was studied with high-voltage-off data, which eliminated the accelerated-proton peak. This component appears to fit well to a double exponential.
\begin{table}[h]
\caption{Averages over all detectors of background fractions for each  background component.} 
\begin{ruledtabular}
\begin{tabular}{lc}
Source &  $\bar\epsilon_b\ (10^{-3})$  \\
\hline
\\ [-8 truept]
Accidental coincidences & $2 \pm 2$\\
Low-energy continuum & $6 \pm 2$  \\
Minimum Ionizing & $0.003 \pm 0.002$ \\
\end{tabular}
\end{ruledtabular}
\label{tb:Backgrounds}
\end{table}

\subsection*{Background-related contributions to $\tilde D$}

The asymmetries $w^{p_ie_j}$ must be corrected to determine the background-free asymmetries $w_0^{p_ie_j}$. The correction is
\begin{equation}
w_0^{p_ie_j} -w^{p_ie_j}=\epsilon^{p_i e_j}_{b}w^{p_i e_j}-\epsilon^{p_i e_j}_{b}w_{b}^{p_ie_j}.
\label{eq:BackgroundCorrection}
\end{equation}
The  first term is the multiplicative correction to $w^{p_i e_j}$ due to the dilution of the asymmetries by backgrounds. This dilution can produce a contribution if the background corrections are non-uniform and do not completely cancel when the $w^{p_ie_j}$ are combined into  $\tilde D$.  The multiplicative correction to $\tilde D$ was determined using measured $\epsilon^{p_ie_j}$ and $w^{p_ie_j}$ according to Eq.~\ref{eq:BackgroundCorrection} resulting in a correction to $\tilde D$ of $(0.03\pm 0.09)\times 10^{-4}$. The multiplicative correction is dominated by the exponential components of the prompt events shown in Fig.~\ref{fg:PromptSpectrum}. The second term in Eq.~\ref{eq:BackgroundCorrection} is due to a possible asymmetry in the background, $w_b^{p_ie_j}$. The $w_b^{p_ie_j}$ were found to be uniform and consistent with zero across the detector and were combined according to Eq.~\ref{eq:Dtilde}. The resulting $\tilde D_b$ was scaled by the sum of $\bar\epsilon_b$ for all contributions to determine the additive correction to $\tilde D$ of $(-0.07\pm 0.07)\times 10^{-4}$. For the result reported in reference~\cite{rf:emiTPRL}, no correction was made for the additive-background effect.

\subsection{Backscattering}
\label{sec:Backscattering}

Backscattering events fall into two categories: 1)  particles scattered from somewhere in the apparatus to the proton or electron detector, and 2) incident protons and electrons  that scatter from the respective detectors  without registering a hit. 
Particles backscattered from the proton and electron detectors  lead to multiplicative and additive corrections to the $w^{p_i}$ that can affect $\tilde D$ in a manner similar to background effects. Backscattering also affects $K_D^{p_i}$ and is discussed in section~\ref{sec:KDerror}.
 
\begin{figure}
\centerline{\hskip -1.85 truein\includegraphics[width=5in]{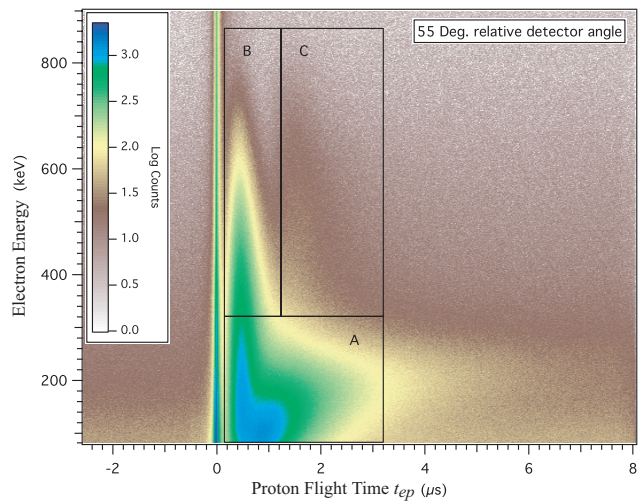}}
\caption{Data for electron energy~vs.~proton-electron delaytime for proton cell $p_1$ in coincidence with electron detector $e_1$ at $55^\circ$. The boxes indicate regions of kinematically allowed proton-electron coincidences (region A), electron backscattering from detector elements across from the proton cell (region B), and proton backscattering (region C).  }
\label{fig:MCTOF1}
\end{figure}

Backscattering corrections to $\tilde D$ were determined by Monte Carlo simulations with empirical validation based on studies of the $35^\circ$ and $55^\circ$  proton-electron coincidences. 
Backscattering probabilities are similar for all proton-electron-detector combinations; however, the rate of neutron-decay coincidence counts for the $35^\circ$ and $55^\circ$  pairings is about a factor of ten less than for the $125^\circ$ and $145^\circ$ pairings used to determine $\tilde D$. Data from the $35^\circ$ and $55^\circ$ pairings are shown in the electron-energy-proton-time-of-flight plane in Fig.~\ref{fig:MCTOF1}. The  boxes indicate three distinct kinematic regions: primary neutron-decay events for the $35^\circ$ and $55^\circ$ proton-electron coincidences are restricted to a region included in region A; region B consists of higher energy electrons, which are primarily electrons with initial momentum directed at large angles with respect to the detected proton but detected in the $35^\circ$ or $55^\circ$ electron detector due to backscattering; region C corresponds to high energy electrons from events with large proton-electron angular separation and delayed proton time-of-flight, which are primarily due to proton backscattering. The backscatter fractions are estimated from the data by integrating within the boxes in Fig.~\ref{fig:MCTOF1} or by fitting the time-of-flight spectra and  are consistent with the Monte Carlo predictions.

\subsubsection*{Electron backscattering}
\label{sec:electronbackscattering}

Electron backscattering was incorporated into the Monte Carlo simulations using {\scshape penelope}. The multiplicative correction to $\tilde D$ was determined using  estimates of the backscattering fractions from  the Monte Carlo to correct the measured $w^{p_ie_j}$ similarly to the background corrections given in Eq.~\ref{eq:BackgroundCorrection}. 
 With the correction  applied, $\tilde D$ changes by \mbox{$(0.11\pm 0.03)\times 10^{-4}$,} where the
 uncertainty reflects the 20\%  uncertainty assigned to the Monte-Carlo results  due to  limitations of the detector and beam model and limited knowledge of backscattering at these low energies. The additive correction would vanish for isotropic backscattering and uniform detector efficiency; however, the beam expansion, magnetic field, and detector elements break these symmetries. The result for the additive correction to $\tilde D$ is \mbox{$(0.09\pm 0.07)\times 10^{-4}$,} and is limited by the statistical precision of the Monte Carlo due to the small fraction of backscatter events. This  additive-electron-backscattering correction was not applied in the analysis reported in reference~\cite{rf:emiTPRL}.

\subsubsection*{Proton backscattering}
\label{sec:protonbackscattering}

Backscattered protons in principle affect the asymmetries similarly to
backscattered electrons. Proton  backscattering probabilities determined with {\scshape SRIM} were input into the Monte Carlo.  While the proton backscattering probabilities at the energies characteristic of neutron decay are comparable to electron backscattering probabilities, there is a large probability for neutralization ~\cite{rf:Allison58}. Detailed calculations lead to the conclusion that proton-backscattering effects on $\tilde D$  are approximately an order of
magnitude smaller than electron-backscattering effects, and we set an upper limit on the correction to $\tilde D$ of less than \mbox{$0.03\times 10^{-4}$}. This upper limit   is reflected in  the uncertainty  given in Table~\ref{tb:Systematics}.

\subsection{Efficiency corrections}

\label{sec:Nonuniformbeta}

\label{sec:CoulterEffect}

The $w^{p_ie_j}$ from Eq.~\ref{eq:ws} depend directly on the energy dependence of the efficiencies through the beta and neutrino correlations. Spatial variations of the proton and electron energy dependence break the symmetry assumed in combining proton-cell data as given in Eq.~\ref{eq:Dtilde}. In this section, we discuss the effects of axial and azimuthal variation of the electron energy thresholds and the variation of the proton-energy-dependent efficiency as a result of changes of individual SBD thresholds.  

\subsubsection{Nonuniform electron-energy  thresholds}

From Eq.~\ref{eq:v_L}, it can be seen that the proton-electron correlation leads to a difference of $\beta_e$ for the two electron detectors that pair with $p_1$  ({\it e.g} $e_3$ and $e_2$). This leads to a contribution to $v^{p_i}$ from the beta-asymmetry ($A$-coefficient correlation), which cancels for a uniform neutron beam  and detector when the $v^{p_i}$ from two axially-symmetric proton cells are averaged ({\it e.g.}~cells $p_1$  and $p_{15}$). In the case of nonuniform electron-detector efficiencies, specifically due to spatial variation of the electron-energy threshold, this cancellation is not perfect and leads to a dependence on the azimuthal proton-cell position, {\it i.e.} a dependence on proton-detector plane; however, another symmetry arises because each electron detector is paired with two azimuthally opposed proton cells and contributes to the two $v^{p_i}$ with opposite sign. This results in cancellation of any efficiency variation to first-order in $\sin\phi^{p_i}$, as given in Eq.~\ref{eq:v_T}. The $\sin2\phi^{p_i}$ dependence of $v^{p_i}$, however,  does not cancel.

\begin{figure}
\centerline{\includegraphics[width=3.5in]{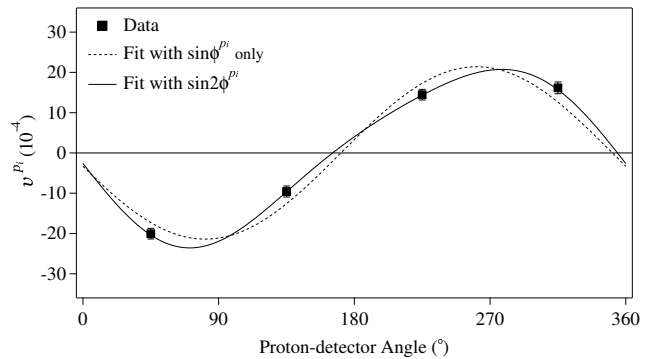}}
\caption{Results for $v^{pi}$ averaged for each of the four proton-detector planes. The dashed line is a fit to $\sin\phi^{p_i}$, and the solid line includes a $\sin2\phi^{p_i}$ term.  Monte-Carlo simulations show that a $\sin2\phi^{p_i}$ term will arise from $\approx$ 10 keV differences of the electron-detection thresholds for upstream and downstream ends of the electron detectors.
\label{fg:NonUniformBetavs}}
\end{figure}

The electron energy thresholds were measured to differ between upstream and downstream phototubes by as much as 10-20 keV resulting in the $\sin2\phi^{p_i}$ dependence evident in Fig.~\ref{fg:NonUniformBetavs}, where the combined data for $v ^{p_i}$ for all proton cells in a single detector plane are shown  as a function  of $\phi^{p_i}$. The dashed line shows the best fit to $\sin\phi^{p_i}$ only, and the solid curve includes a $\sin2\phi^{p_i}$ contribution.  Monte-Carlo studies with an upstream-downstream variation of electron threshold of approximately 10 keV for a single electron-detector show a similar $\sin2\phi^{p_i}$ contribution.

The Monte Carlo also shows that the axial dependence of the electron thresholds are cancelled when all four proton-detector planes are combined and therefore, does not contribute significantly to $\tilde D$  as long as the proton efficiencies are energy independent.  The variation of proton energy dependent efficiency, discussed below, could couple with an electron-threshold variation. Coupling between the observed relative efficiencies of proton detection and the Monte-Carlo model of electron threshold variations had an effect on $\tilde D$ of $(-0.04\pm 0.1)\times 10^{-4}$. 

\subsubsection{Proton-SBD efficiency variations (Shift-threshold effect)}

\begin{figure}
\includegraphics[width=3.5 truein]{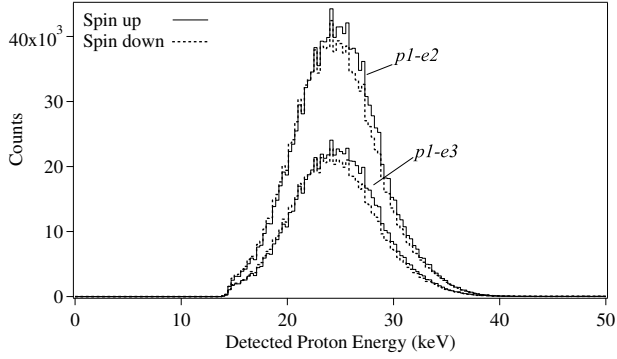}
\caption{Proton-cell SBD spectra for a single proton detector (e.g.~$p_1$) for both electron detectors $e_2$  and $e_3$ for both spin-flipper states: up (neutron polarization parallel to $\vec B$) and down (neutron polarization anti-parallel to $\vec B$). }
\label{fg:ProtonSpectrum}
\end{figure}

\begin{figure}
\includegraphics[width=3.5in]{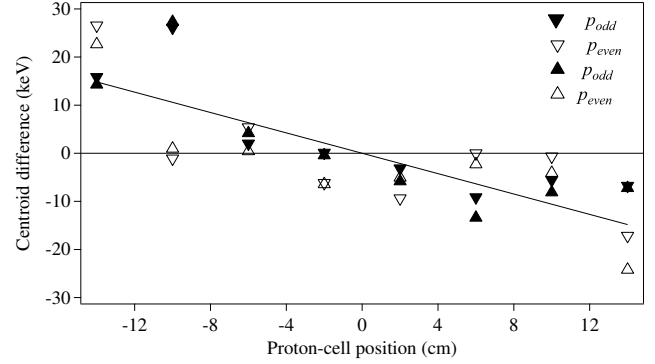}
\caption{Differences of proton-detector centroids for proton cells $p_1$-$p_{16}$ as a function of axial position for both 125$^\circ$ and 145$^\circ$ proton-electron-detector pairs. Fluctuations from cell-to-cell reflect variations of energy loss and other features of individual proton cells. Odd and even proton detector refer to $p_1$, $p_3$, {\it etc.} and $p_2$, $p_4$, {\it etc.}, respectively.
}
\label{fg:ProtonCentroids}
\end{figure}

A set of typical calibrated proton-cell SBD spectra is shown in Fig.~\ref{fg:ProtonSpectrum} for both spin-flipper states and for both 125$^\circ$ and 145$^\circ$ coincidence angles. The electron detector at $145^\circ$ has approximately twice the rate as the $125^\circ$ detector due to the proton-electron  correlation, which  favors  back-to-back proton and electron momenta. The spectra for the two spin-flipper states differ in both area (total number of counts, $N_\pm^{p_ie_j}$) and the spectral distribution due to the beta asymmetry, which affects the angular distribution of electrons and thus the proton momentum. The energy deposited in the SBD in turn depends on the incident proton momentum, position and angle of incidence on the proton cell. We characterize each spectrum by a centroid, which depends on the neutron spin sate. The difference of the centroids for the two spin-flipper states varies with the axial position of each proton cell and varies in magnitude from zero to about 125 eV. In Fig.~\ref{fg:ProtonCentroids}, we show the difference of the  centroids as a function of proton-cell position for a single proton-detector plane ($p_1$-$p_{16}$) paired with the electron detectors at 125$^\circ$ and 145$^\circ$.
 
\begin{figure}
\centerline{\includegraphics[width=3.75in]{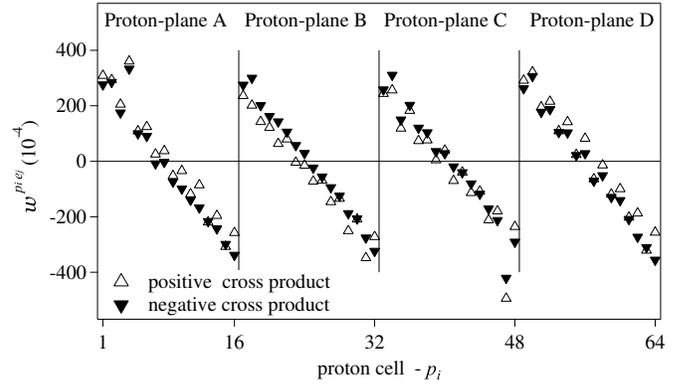}}
\caption{$w^{p_ie_j}$ for all proton cells paired with electron detectors at $\pm 125^\circ$ and $145^\circ$ with respect to the proton plane. The  open triangles pointing up are for positive cross product $\vec p_p\times \vec p_e$, and the filled triangles pointing down are for negative cross product. The $v^{p_i}$ shown in Fig.~\ref{fg:vs} are found from the difference of the two $w^{p_ie_j}$ for each proton cell. }
\label{fg:ws}
\end{figure}

\begin{figure}
\centerline{\includegraphics[width=3.75in]{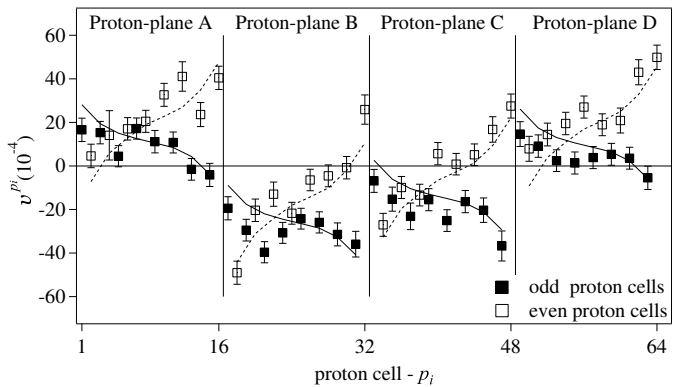}}
\caption{$v^{p_i}$ for all proton cells for the experiment. The solid and dashed lines are from the Monte Carlo simulation for realistic experimental conditions shown in Fig.~\ref{fg:MCvbysliceB5p6Gauss2} with adjustment for each proton-cell plane to account for transverse polarization and beam-shape effects.}
\label{fg:vs}
\end{figure}

For the SBD spectra shown in Fig.~\ref{fg:ProtonSpectrum}, the low energy portion of the spectrum shows that the applied threshold allows detection of almost all of the protons for both electron detectors and both spin-flipper states. The SBD  thresholds were adjusted throughout the run in order to maintain the electronics' noise at a manageable level and minimize dead-time losses. In some cases, the  threshold cut significantly into the SBD spectra introducing a proton-energy-dependent detection efficiency. Due to the dependence on neutron polarization, this energy dependence could lead to a significant effect on the $w^{p_ie_j}$ for an SBD with a high threshold, though the effect on $v_z^{p_i}$ is largely mitigated  because the low-energy portion of the proton energy spectrum is nearly equally affected for the two electron detectors. 
Fig.~\ref{fg:ws} shows that there are several notable anomalies of the $w^{p_i e_j}$, for example for proton cells $p_4$  and $p_{12}$; however, as expected, the anomalies are similar for both electron-detector pairings. Thus the $v^{p_i}$,  as shown in Fig.~\ref{fg:vs}, do not appear to fluctuate significantly compared to other proton cells. 

To estimate the effect on $\tilde D$, neutron-polarization dependent proton energy spectra were generated by Monte Carlo for all proton-detector-electron-detector pairs and convoluted with model detector response. The model detector response (modified gaussians with separate widths above and below the centroid) were  based on measured proton-SBD spectra. These spectra and the thresholds  varied significantly during the course of the experiment, and the set of response functions was therefore based on an average over the data subsets.   Parameters of the response functions  were varied over a range characteristic of the variations during the experiment in order to estimate the uncertainties. The estimated correction to $\tilde D$ is $(-0.29\pm 0.41)\times 10^{-4}$.
The effect was also estimated by correcting $w^{p_ie_j}$ on a day-by-day basis using the centroid shift and slope at threshold with consistent results.

\subsection{ Beam Expansion Correction}

\label{sec:ExpandingBeamCorrection}

\begin{figure}
\begin{centering}
\centerline{\includegraphics[width=3.5 truein]{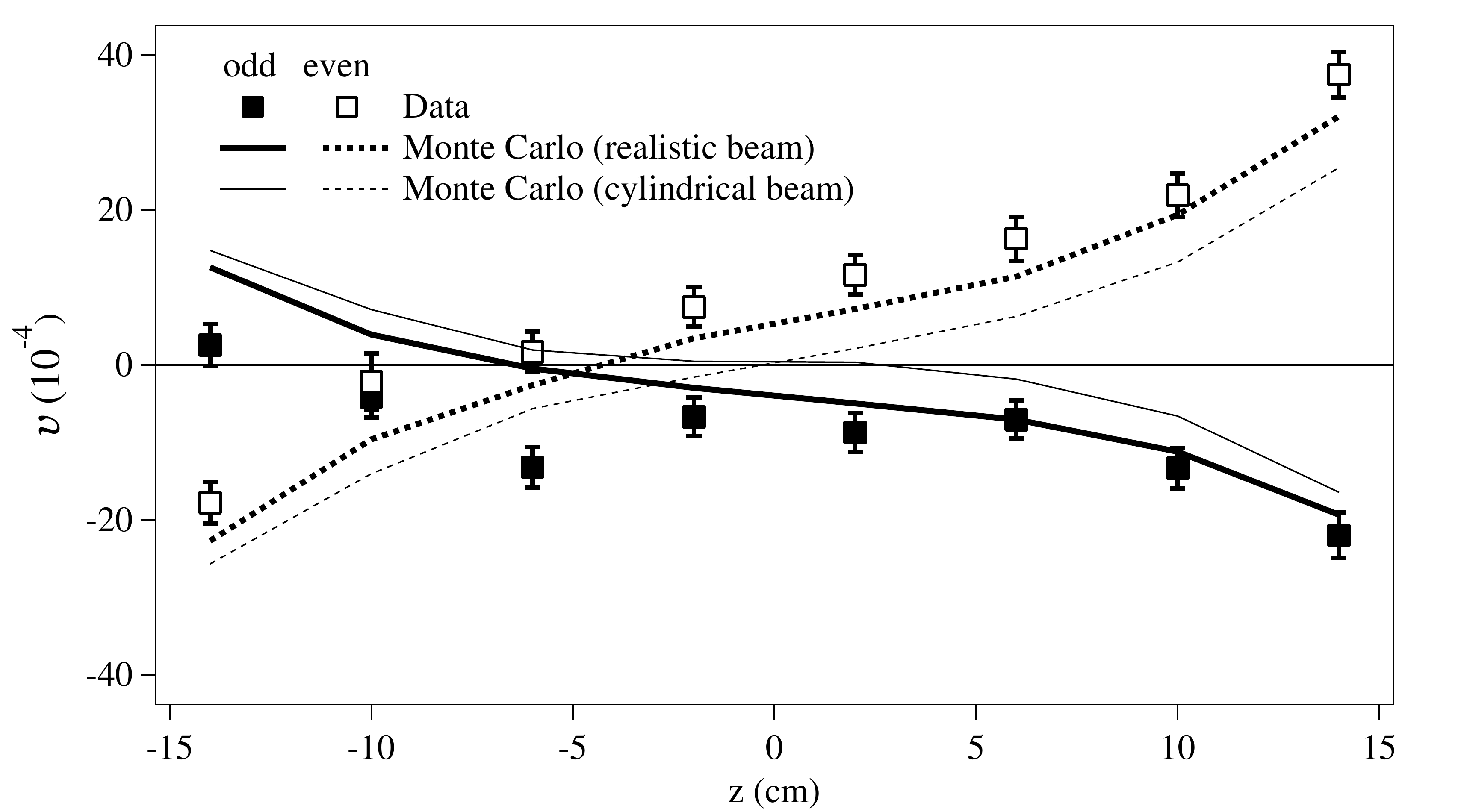}}
\caption{The broken symmetries due to magnetic field and beam expansion contribute to a  $\tilde D$ for longitudinal polarization. The 560~$\mu$T magnetic field makes the magnitude of $v^{p_i}$ larger for even detectors than for odd detectors, {\it i.e.} the dependence of $v^{p_i}$ vs. $z^{p_i}$ differs  largely due to the difference of $\langle\beta_e^{p_i p_j}\rangle$ for the two electron detectors, {\it e.g.}
the expanding beam makes the magnitude of $(\langle\beta_e^{p_1 e_3}\rangle-\langle\beta_e^{p_1e_2}\rangle)$ larger for downstream detectors since the larger beam leads to larger proton-electron angular separation.
\label{fg:MCvbysliceB5p6Gauss2}}
 \end{centering}
 \end{figure}

 The neutron beam  expansion affects the cancellation of the beta-asymmetry (Eq.~\ref{eq:v_L}) for axially-paired proton cells because the average of the electron velocity $\langle\beta_e\rangle$  depends on the radial size of the beam: for decay vertices further from the center of the detector, the proton-electron angular separation is larger, which corresponds to higher energy electrons.  This would be largely cancelled by the combination  of the $v^{p_i}$ prescribed by Eq.~\ref{eq:Dtilde} because the difference of $\langle\beta_e\rangle$ enters with opposite sign for two adjacent proton cells ($p_1$ and $p_2$); however, the magnetic field also affects the electron-proton angular correlation and therefore $\langle\beta_e\rangle$ resulting in a magnetic-field dependence. The 560~$\mu$T magnetic field makes the magnitude of the difference for the two electron detectors ({\it e.g.}~$p_1 e_3$ and $p_1 e_2$) smaller for odd detectors ($p_1$, $p_3$, $p_5$, {\it etc.}) than for even detectors ($p_2$, $p_4$, $p_6$, {\it etc.}). The result is that the slopes of $v^{p_i}$ vs. $z^{p_i}$ have different magnitudes for even and odd proton cells.
Fig.~\ref{fg:MCvbysliceB5p6Gauss2} shows a comparison of a uniform, nonexpanding neutron beam and the realistic beam based on the beam-distribution maps of Fig.~\ref{fg:BeamMaps}. A possible systematic effect arising from the beam expansion coupling with a polarization gradient is discussed in section~\ref{sec:Polnonuniformity}.

\begin{figure}
\begin{centering}
\centerline{\includegraphics[width=3.25 truein]{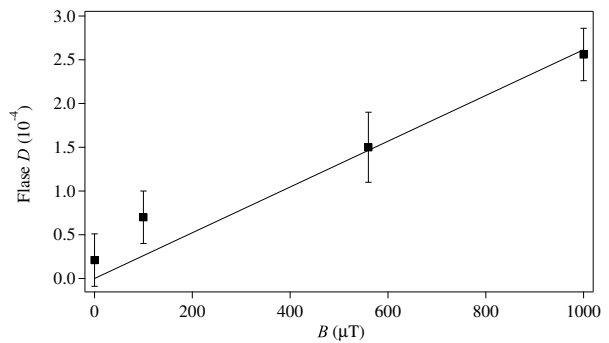}}
\caption{The beam expansion effect as a function of magnetic field determined by Monte-Carlo simulations using the measured beam profiles. The solid line is a fit to the slope, constraining the effect to zero at $B=0$. The error bars represent the statistical uncertainty of the Monte-Carlo.
\label{fg:DexpvsB}}
 \end{centering}
 \end{figure}

To estimate the correction, Monte Carlo simulations were run using the measured neutron-beam distribution. A number of simulations were run with different beam distributions produced by shifting the maps by up to 2 mm and rotating by 5$^\circ$, both significantly greater than the mechanical alignment constraints on the beam-line and detector components. In Fig.~\ref{fg:DexpvsB} we show the results for the beam expansion effect as a function of magnetic field.
 The  estimated correction at $B=560\ \mu$T is  $\tilde D$ is $(-1.5\pm 0.4)\times 10^{-4}$, where the uncertainty is mainly due to the change of the effect as the beam-distribution models were varied.  

\subsection{Transverse polarization effects}
\label{sec:ATP}
\label{sec:Polnonuniformity}

As discussed in section~\ref{sec:IdealExperiment},  transverse asymmetries arise in the experiment  due to  the polarization misalignment with respect to the detector axis. The misalignment  was determined from the data combined with transverse-polarization calibration runs combined with Monte-Carlo simulation of the beam-shape effects and by magnetic field maps measured before and after the run. The transverse polarization leads to a contribution $v_T^{p_i}$ given in Eq.~\ref{eq:v_T}. 
For uniform-symmetric beam and  polarization, $v_T^{p_i}\propto\sin(\phi_P-\phi^{p_i})$ and is cancelled when data from two opposing proton cells  are combined; however, due to the asymmetric
emiT-II beam, the cancellation is not complete. Displacement of the beam perpendicular to the transverse polarization would result in a contribution to $\tilde D$  that is the product of the polarization misalignment  and the {\em perpendicular} misalignment of the beam.  This is called the ATP or asymmetric-beam-transverse-polarization effect. An additional effect arises if the transverse polarization is nonuniform, specifically that $\sin\phi_P$ varies along the axis of the detector.

\subsubsection{Asymmetric Beam/Transverse Polarization Effect}

\begin{figure}
\begin{centering}
\includegraphics[width=3.25 truein]{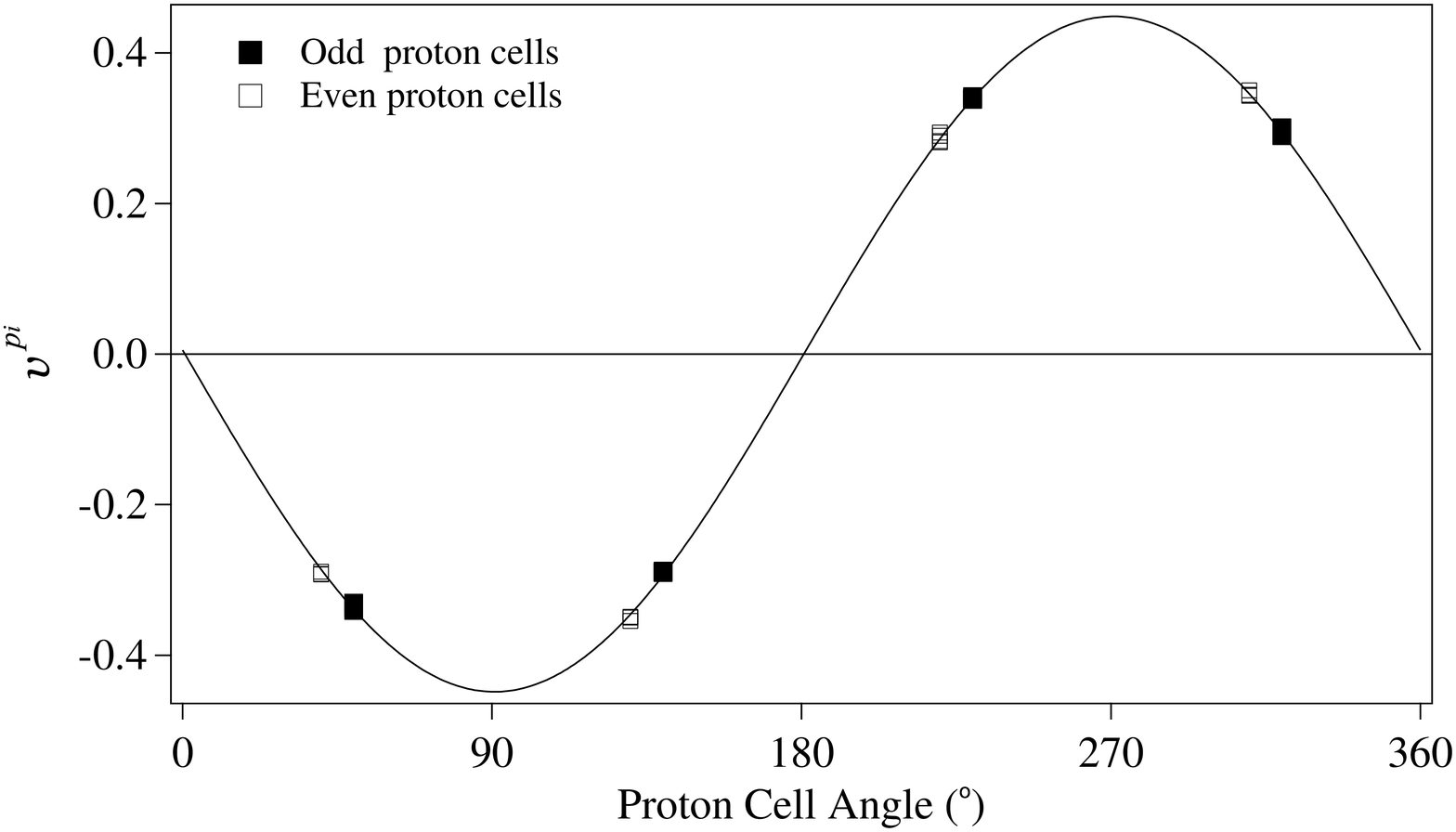}
\caption{Results of a transverse-polarization calibration run with $\phi_P=0^\circ$, {\it i.e.} polarization pointing in the $\hat x$ direction (see Fig.~\ref{fg:emiTDetector} for coordinate system definitions).  Each point combines data for two proton cells with the same $|z^\alpha|$. The solid line is a fit to the form of Eq.~\ref{eq:v_T}. The average amplitude for 11 calibration runs is $<v_{T}>=0.4565\pm 0.0128$, where the uncertainty is the standard deviation of the 11 amplitudes.
\label{fg:ATPRun}}
\end{centering}
\end{figure}

We have studied the ATP  effect using transverse polarization calibration runs.  Data from a transverse-polarization calibration run with $(\theta_P,\phi_P)=(90^\circ,180^\circ)$ is shown in Fig.~\ref{fg:ATPRun}. This amplifies the ATP effect, which is then scaled by the size of the transverse polarization for the run to determine the effect on $\tilde D$.
Calibration runs at several  azimuthal angles $\phi_P$ map out the effect to provide a more accurate estimate. The size of the effect was also estimated in Monte Carlo simulations with $\theta_P=90^\circ$ and several different values of $\phi_P$. 

For calibration runs, the axial magnetic-field coils were used to cancel the Earth's field component along the detector axis, and the transverse-field coils were used to produce a field of about 100 $\mu$T perpendicular to the detector axis. The azimuthal angle could be selected; however, due to power supply limitations, the magnitude of the transverse field could not be maintained at 100 $\mu$T for all azimuthal angles. In Fig.~\ref{fg:ATPRun}, we show  the average $v^{p_i}_T$ for even and odd proton cells as a function of $\phi^{p_i}$ from a transverse polarization calibration run with $\theta_P=90^\circ$ and $\phi_P=0^\circ$.
A total of eleven polarization-calibration runs were taken over the course of the experiment. For each run,  the amplitude and the offset are determined by fitting the data to a sinusoid. The average amplitude for all calibration runs with $\theta_P=90^\circ$ is 
\begin{equation}
<v_{T}>=0.456\pm 0.013,
\label{eq:vTResult}
\end{equation}
where the uncertainty is the standard deviation  of the eleven amplitudes. 

For the data runs, $\theta_P$ is small; thus the asymmetric beam shape contributes significantly to the $\phi^{p_i}$ dependence of the $v$s.  The beam maps shown in Fig.~\ref{fg:BeamMaps} show that, in addition to the beam expansion, the center of gravity of the beam is horizontally displaced from the detector axis and rises in the vertical direction for the downstream map. 
To separate the beam-shape and  transverse-polarization effects, the $v^{p_i}$ were corrected for the beam-shape effect, which was determined from Monte Carlo simulations with no transverse polarization. The value of $\sin\theta_P$ is found by dividing the amplitude of a sinusoid fit to the corrected $v^{p_i}$ by the average amplitude from the transverse-polarization calibration runs.  In Fig.~\ref{fg:DataRun}, we  show results of Monte-Carlo studies for the beam-shape along with data and the resulting estimate of the transverse polarization effect. The Monte Carlo results for the beam-shape effect are very sensitive to the small differences in the beam-map registration and orientation. This results in large uncertainties in the transverse  polarization direction. 
The results are
$$
\sin\theta_P= (8.5\pm 4.3)\times 10^{-3}
\quad
\phi_P=(40\pm 72)^\circ.
$$
The uncertainty is determined from the uncertainty on $\langle v_T\rangle$ and  errors arising from the beam-shape correction and the $\sin2\phi_P$ effect 
(Fig.~\ref{fg:NonUniformBetavs}). Alternatively, using the magnetic-field maps discussed in section~\ref{sec:Bfieldmap},
 we estimate
$$
\sin\theta_P= (3.8\pm 1.0)\times 10^{-3}
\quad
\phi_P=(45\pm 5)^\circ.
$$
which is consistent with the analysis using the data and beam-shape correction. To determine the ATP correction, we use the 1-$\sigma$ limits from the transverse-polarization calibration runs, which give 
$$\theta_P\le 12.8\times 10^{-3}\quad   -32^\circ \le \phi_P \le 112.$$ 
These limits provide more conservative bounds on the ATP effect than those base on the magnetic-field maps.

In Fig.~\ref{fg:DATPNewBeamMapCorr}, we show  the data with fit  and the result for the Monte-Carlo model of the experiment for $\sin\theta_P=8.5\times 10^{-3}$.  From the fit and the values of $\theta_P$ and $\phi_P$ given above, we determine the correction to $\tilde D$ of {\mbox{$(-0.07\pm 0.72)\times 10^{-4}$.} The relatively large uncertainty is due to the uncertain azimuthal orientation of the polarization.

\begin{figure}
\begin{centering}
\includegraphics[width=3.5 truein]{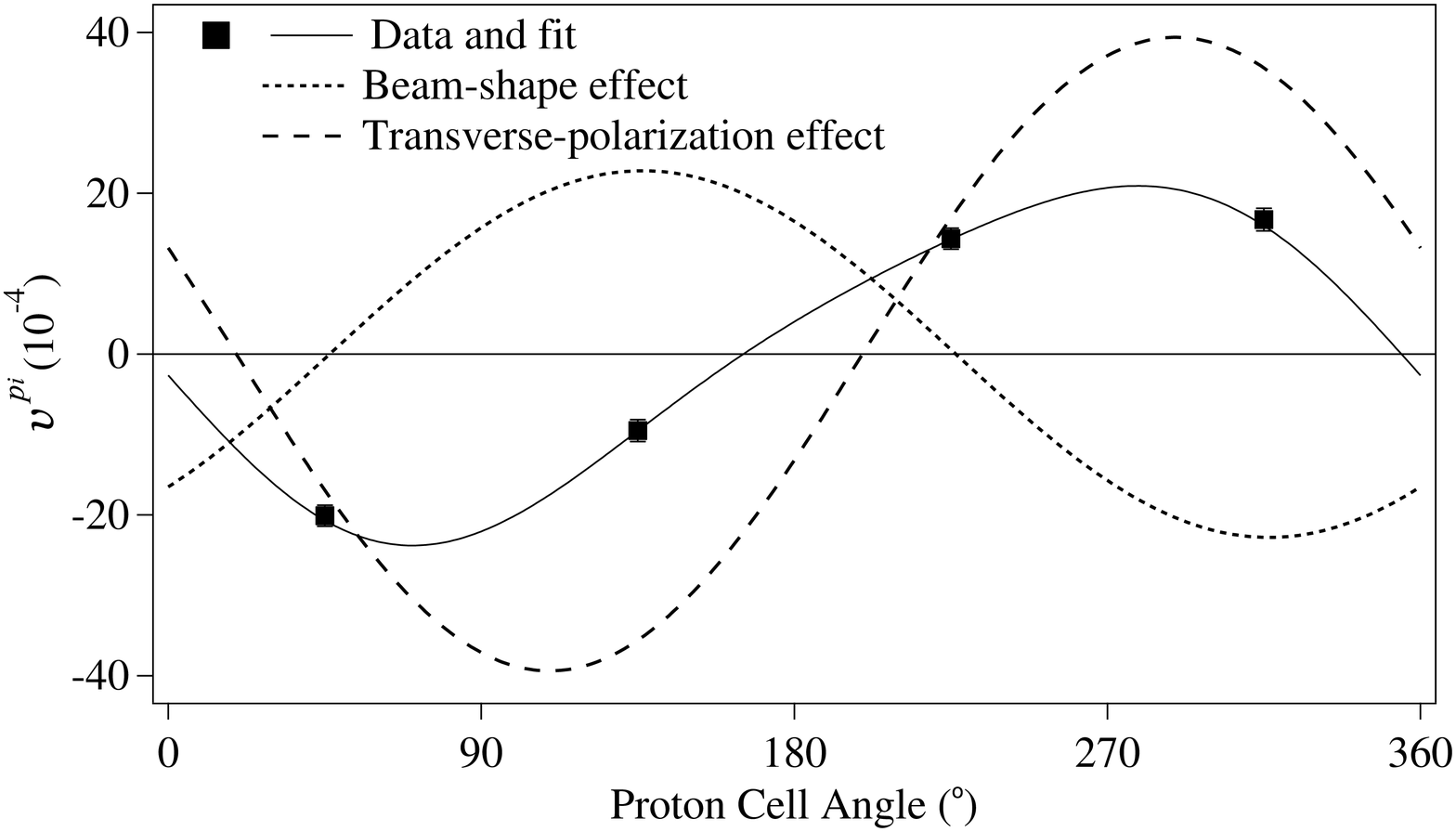}
\caption{Beam-shape and transverse-polarization effects on $v^{p_i}$.  The filled squares show the average $v^{p_i}$ for each proton-detector plane as a function of azimuthal angle for the experiment. The solid line is a  fit to the data including the $\sin2\phi^{p_i}$ contribution. The dotted and dashed lines are, respectively, the estimated effects of the beam shape  and the transverse polarization. The beam shape effect was determined by Monte Carlo simulations for the beam maps shown in Fig.~\ref{fg:BeamMaps}; the transverse polarization was estimated from the difference of the $\sin\phi^{p_i}$-only fit (Fig. ~\ref{fg:NonUniformBetavs}) and the beam-shape effect. 
\label{fg:DataRun}}
\end{centering}
\end{figure}

\begin{figure}
\begin{centering}
\centerline{\includegraphics[width=3.5 truein]{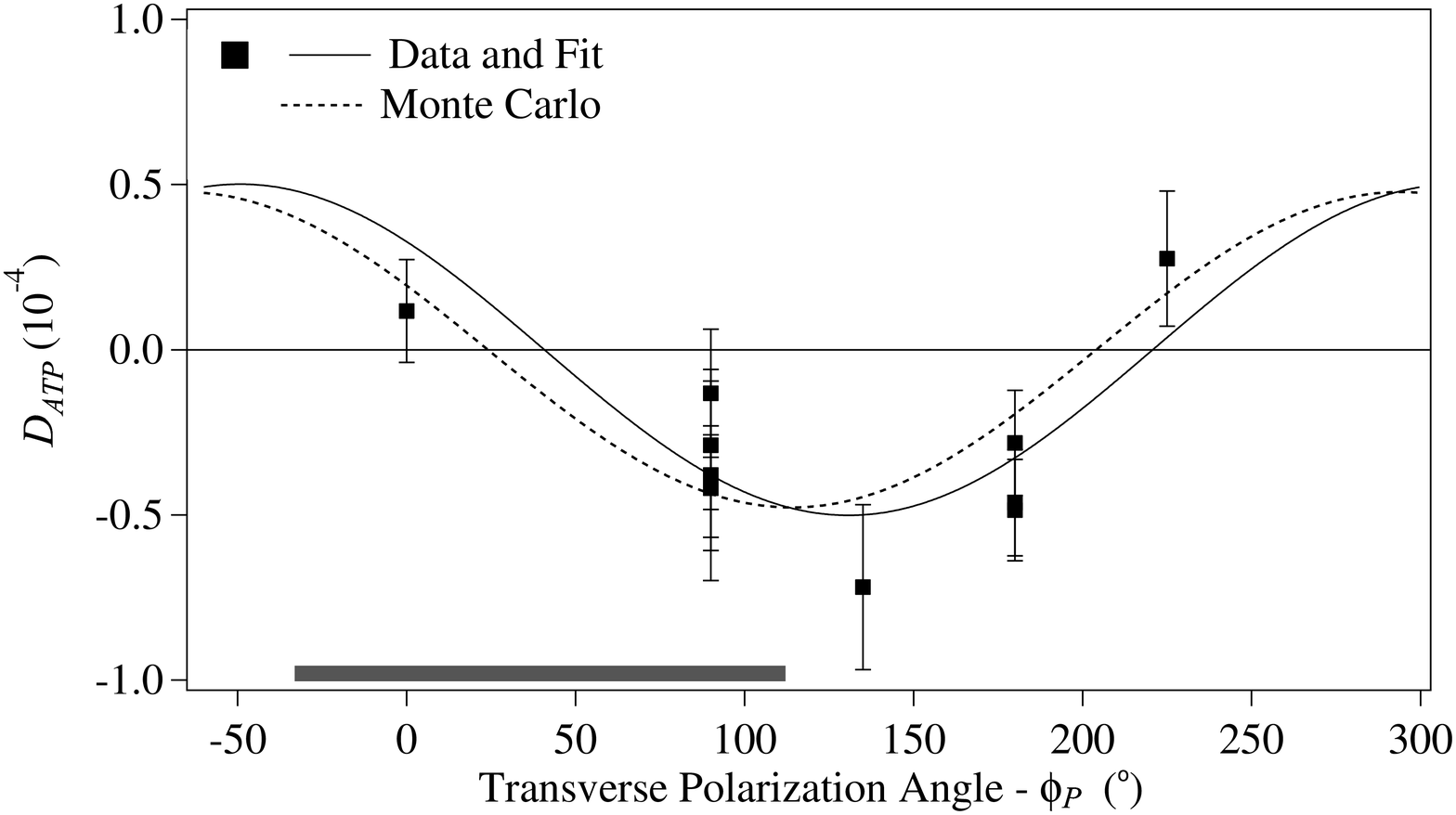}}
\caption{ATP effect on  $\tilde D$ as a function of $\phi_P$. The data points were determined from the eleven polarization-calibration runs scaled by $\sin\theta_P=8.5\times 10^{-3}$. The solid line is a sinusoidal fit to the data with the offset constrained to be zero.  The dashed line, given by $0.48\times 10^{-4}\sin(\phi_P-204^\circ)$, is the Monte-Carlo prediction based on beam maps and $\sin\theta_P=8.5\times 10^{-3}$. The horizontal bar extending from -33.1$^\circ$ to 112.1$^\circ$ shows the range of $\phi_P$ based on Monte Carlo estimates using reasonable variations of the beam-shape map registration. 
\label{fg:DATPNewBeamMapCorr}}
 \end{centering}
 \end{figure}

\subsubsection{Polarization twist}

\noindent
A twist of the polarization, that is an upstream-downstream difference in $\phi_P$, would not be cancelled in combining 16 proton cell $v^{p_i}$ into $\tilde D$.  
 A limit on the polarization twist was estimated by separating data  for upstream \mbox{($z^{p_i}\le -6$ cm)} and downstream \mbox{($z^{p_i}\ge 6$ cm)} proton cells and correcting for beam-shape effects. The beam-shape corrected data are shown shown in Fig.~\ref{fg:PolTwistBeamShapeMC}. Sinusoid fits provide estimates for $\phi_P$ for upstream and downstream portions of the detector of $\phi_P^{up}=14.3^\circ\pm 5.2^\circ$ and $\phi_P^{dn}=16.6^\circ\pm 6.0^\circ$, from which we use to estimate a maximum upstream-downstream difference $\Delta\phi_P=13.5^\circ$. We double  $\Delta\phi_P$ to account for a change over the entire length of the proton-detector plane and use the fits from Fig.~\ref{fg:DATPNewBeamMapCorr} to set an upper limit on the magnitude of the  ATP-twist effect of  $\tilde D$ of $0.24\times 10^{-4}$. This upper limit is reflected in the uncertainty listed in Table~\ref{tb:Systematics}.
\begin{figure}
\begin{centering}
\centerline{\includegraphics[width=3.5 truein]{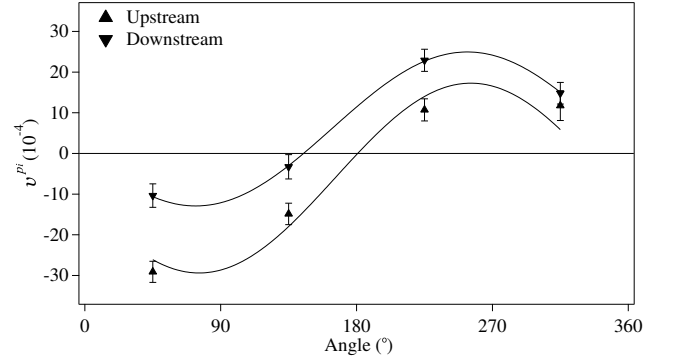}}
\caption{Data, corrected for beam-shape effects, for $v^{p_i}$ for upstream and downstream proton cells. The solid lines indicate sinusoidal fits with $\phi_P^{up}=14.3^\circ\pm 5.2^\circ$ and $\phi_P^{dn}=16.6^\circ\pm 6.0^\circ$.
\label{fg:PolTwistBeamShapeMC}}
 \end{centering}
 \end{figure}

\subsubsection{Polarization nonuniformity}

A nonuniform neutron polarization affects the cancellation of the electron ($A$) and antineutrino ($B$) asymmetries in $\tilde D$. The polarization map, which measures the combination of neutron-beam polarization $P$ and the  analyzer power $A_P$, is shown in Fig.~\ref{fg:PolMap}. Assuming $P=A_P$,  the polarization as a function of position was convolved with the density of neutron-decay vertices determined by Monte Carlo to determine an effective polarization for the decays detected by each proton cell. This was also studied using data from both transverse-polarization calibration runs and normal running. This effective polarization varied by a few percent along the beam axis due to the beam expansion; however, the average over all cells in each proton-detector varied by less than 0.005.  The effect on $\tilde D$ was estimated by  Monte-Carlo using the averaged polarization for each proton-detector plane.  A possible transverse polarization gradient ($\frac{\partial P}{ \partial x}$ or $\frac{\partial P}{ \partial y}$) was  also investigated with the Monte Carlo and found to have a negligible effect.

\subsection{Uncertainties in Polarization and $K_D$}
\label{sec:KDerror}

Uncertainties in $P$ and $K_D$ lead to errors proportional to $\tilde D$.  The polarization analysis described in section~\ref{sec:Polarization} results in the lower limit  $P>0.91$ (90\% c.l.), and we take $P=0.95\pm 0.05$ for the purpose of analysis. The resulting contribution to the uncertainty on $\tilde D$ is $0.05\tilde D$, which is given in Table~\ref{tb:Systematics}.
The kinematic quantities $K^{p_i}_D$ depend on the proton-electron angular separation due primarily to the cross-product $\vec p_p\times \vec p_e$, and are different for the $125^\circ$ and $145^\circ$ proton-electron-detector pairs. The $K^{p_i}_D$ were determined by Monte Carlo simulations, and a number of parameters were varied   including beam size and shape,  electron backscattering fraction, and electron-energy threshold. For proton-detector-electron-detector pairings at 125$^\circ$ and 145$^\circ$ we find
$|K_D^{125}|=0.420\pm 0.008$ and $|K_D^{145}|=0.335\pm 0.016$,
where the uncertainties estimate the variation of $K_D$ as the Monte Carlo parameters are varied.
The average, used to determine $\tilde D$ (see eq.~\ref{eq:v_L}) is
\begin{equation}
\bar K_D=\frac{1}{2}(|K_D^{125}|+|K_D^{145}|)=0.378\pm 0.019.
\end{equation}
The resulting contribution to the uncertainty is \mbox{$0.05\tilde D=0.04\times 10^{-4}$.}

\subsection{Spin flip and timing errors}

Errors in the $w^{p_ie_j}$ can also arise if the neutron polarization, flux, or counting time in each spin-flipper state is dependent on spin-flipper state. In these cases, the error is approximately proportional to $\tilde D$. For example if the neutron flux or density within the detector is dependent on spin-flipper state, the count rates  would be modified from the such that
$N^{p_i e_j}_\pm=N^{p_i e_j}_{0\pm}(1\pm \frac{\delta_F}{2})$,
where $N^{p_ie_j}_0$ is the count rate for a spin-flip symmetric experiment and $\delta_F$ is the fractional change in the flux.  In the event of a dependence of the flux on the spin-flipper state, the asymmetries would be modified from the spin-flip symmetric asymmetries $w^{p_ie_j}_0$. The measured asymmetry is
\begin{eqnarray}
w^{p_ie_j} \approx w_0^{p_ie_j}+\delta_F - w_0^{p_ie_j}\delta_F^2 + \delta_F (w_0^{p_ie_j})^2.
\end{eqnarray}
When the differences of the $w^{p_ie_j}$ are combined into the $v^{p_i}$ and finally $\tilde D$, the correction is 
\begin{equation}
\tilde D_0-\tilde D\approx \delta_F^2 \tilde D - {\cal O}(\tilde D^2),
\end{equation}
where we assume the measured $\tilde D$ is approximately equal to the corrected $\tilde D_0$.
A spin-flip correlated counting time difference ($\delta_T$) and polarization difference ($\delta_P$) introduce similar corrections.

Measured limits on the flux  and counting time correlations with spin-flipper state are $\delta_F<0.004$, and $\delta_T<10^{-8}$ respectively. Thus the corrections to $\tilde D$ due to spin-flip correlated flux and counting time variations are less than $0.01\times 10^{-4}$. The spin-flipper efficiency was estimated to be greater than 0.95, implying $\delta_P<0.05$ and a correction to $\tilde D$ less than $0.04\times 10^{-4}$.

\subsection{Summary of Systematic Error Corrections}

The corrections to $\tilde D$ are summarized in Table~\ref{tb:Systematics}. The polarization and instrumental constant $P$ and $K_D$  are included in the definition of $\tilde D$ and are not included as corrections; however, the uncertainties on both are included in the table. The total correction is the sum of all corrections. The uncertainties are independent and are therefore combined in quadrature to determine the total systematic-error uncertainty.

\begin{table}[htpb]
\caption{Systematic error corrections in units of $10^{-4}$. We determined upper limits on the magnitude of corrections for proton backscattering, polarization non-uniformity, and the ATP-twist effect, and thus these corrections are indicated as 0 with the upper limit indicated by the uncertainty. Corrections for spin-correlated flux and spin-correlated polarization are less than $0.01\times 10^{-4}$, thus no correction was made, and the contribution to the uncertainty was negligible. \\
 $^{\rm a}$ Polarization and $K_D$ are included in the definition of $\tilde D$. \\$^{\rm b}$ Assumes polarization uncertainty of 0.05.}
\begin{ruledtabular}
\begin{tabular}{lcc}
Source 							& Correction							& Uncertainty 	\\
\hline
\\ [-8 truept]
Background additive				 	& -0.07						& 0.07		\\
\quad Multiplicative$^{*}$				&\ 0.03						& 0.09		\\
Electron backscattering additive  		&\  0.09						& 0.07		\\
\quad Multiplicative					&\ 0.11						& 0.03		\\
Proton backscattering 				&\ 0							& 0.03		\\
Electron threshold non-uniformity		&\ 0.04						& 0.10		\\
Proton-threshold effect				&-0.29 						& 0.41		\\
Beam expansion					&-1.50						& 0.40		\\
Pol. non-uniformity					&\   0						& 0.10		\\
ATP - misalignment 					& -0.07						& 0.72		\\
ATP - Twist 						&\  0							& 0.24		\\
Spin-correlated flux					&\  0							& \!\!\!\!$<\!0.01$		\\
Spin-correlated polarization			&\  0							& \!\!\!\!$<\!0.01$		\\
Polarization						&  \ $^{\rm a}$					&\, 0.04$^{\rm b }$  \\
$K_D$							&  \ $^{\rm a}$					& 0.04	\\
Total corrections					&  -1.66						&0.97	\\
\end{tabular}
\label{tb:Systematics}
\end{ruledtabular}
\\ [1 truept]
$^{*}$ In reference~\cite{rf:emiTPRL} this entry had a typographical error.
\end{table}

\section{Results}
\label{sec:Results}

\subsection{Cross Checks}
\label{sec:CrossChecks}
\label{sec:Cuts}
\label{sec:DataSubsetStudies}

Several cross checks have been performed to validate the analysis and search for systematic errors. All cross checks were performed on blinded data. 
The cross checks fall into three main categories: 1) varying the cuts on experimental parameters, 2) breaking up the data into subsets, and 3) alternative definitions of $\tilde D$ that would be equivalent for an ideal experiment.

\subsubsection{Data Cuts}

The original cuts on experimental parameters listed in section~\ref{sec:DAQMonitors} were established using nominal operating parameters with windows set by typical variations that produced useful data ({\it e.g.}~for the flux-gate magnetometer $\hat z$ component: 5.56 G $\le B_z\le$ 5.62 G). To investigate the effects of the cuts,  the nominal  window was  expanded by a factor of two in most cases ({\it i.e.}  5.53 G $\le B_z\le$ 5.65 G), and the change in $\tilde D$ was noted. For the electron-energy threshold the cut was lowered to 40 keV, which was below the  threshold in some channels, and for electron-detector multiplicity, the cut was changed from one to two.
With the exception of the proton acceleration voltage discussed below, 
$\tilde D$ changed by less than $0.1\times 10^{-4}$ or 5\% of the statistical error.

\subsubsection{Data-subset studies}

Each independent measurement of $\tilde D$ combines $v^{p_i}$ from sixteen proton cells, all of which have relatively high efficiency for proton-electron coincidences; however, during the experiment, individual proton-cell SBDs were turned off  for extended periods due to high leakage currents or noise. Possible variations of the results due to varying experimental conditions were studied by   breaking up the experiment into subsets with roughly equal statistical weight. Subsets were separated by several possible changes including proton-acceleration voltage, number of live SBDs, and changes to the magnetic field prior to transverse-polarization calibration runs. For each of the four $\tilde D$, 16 operating SBDs are required; however, a number of data subsets had one or more sets of proton-cells missing, in which case we report the weighted average of all available full sets of 16 proton-cells. The results for the subsets are shown  in  Fig.~\ref{fg:DSuperSliceSubsets}. A possible correlation of $\tilde D$ with high voltage was revealed in the study of cuts and investigated with data subsets shown in Fig.~\ref{fg:DSuperSliceSubsets}.  Assuming $\tilde D$ is independent of acceleration voltage  results in $\chi^2=10.4$ for 12 degrees of freedom. Allowing a linear dependence of $\tilde D$ with acceleration voltage results in $\chi^2 = 5.6$ for 11 degrees of freedom. The change in $\chi^2$ implies a 2.1-sigma slope.  In addition, the acceleration-voltage dependence of the focusing properties was extensively studied by Monte Carlo simulations, which showed no effect correlating $\tilde D$ with acceleration voltage. We therefore consider the preference for a slope to be an accidental correlation.

\begin{figure}
\includegraphics[width=3.5in]{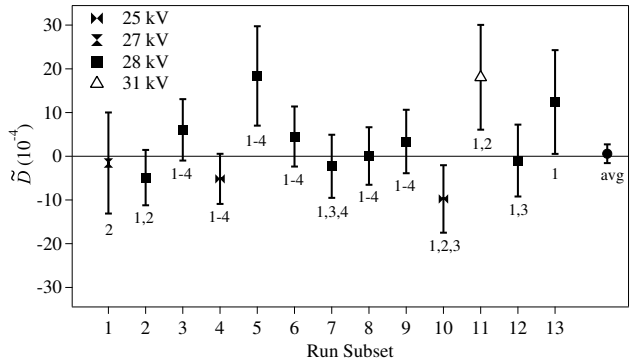}
\caption{Results for $\tilde D$ by data subset. The proton-acceleration voltage and the  proton-cell sets included for each subset are indicated. (Proton-cell set 1 refers to $|z_{p_i}=\pm 2|$ cm, {\it etc}.) The  weighted average of all subsets is $0.58\pm 2.14$ with $\chi^2=10.44$ for 12 degrees of freedom.}
\label{fg:DSuperSliceSubsets}
\end{figure}

\subsubsection{Alternative definitions of $D$}

The definition of $\tilde D$ given in Eq.~\ref{eq:Dtilde} averaged the smallest number of proton cells ({\it i.e.} 16), which cancelled the beta and neutrino asymmetries in the presence of transverse polarization and the 560~$\mu$T magnetic field. Our final result, the weighted average of the four independent determinations of $\tilde D$, thus has the smallest possible statistical error. For a uniform detector and beam, $D$ can also be defined as a) the simple average of all 64 $v^{p_i}$, which differs from $\tilde D$ only due to small changes in weighting of the individual $v^{p_i}$; b) the average of  the combined data from four individual proton-detector planes; c) the offset when the $v^{p_i}$  are fit to a sinusoid (Eq.~\ref{eq:v_T}); d) the offset for the sinusoid fit when the averages of $v^{p_i}$ for each proton-detector plane; or e) the paired proton-cell approach used in the analysis of emiT-I data~\cite{rf:LisingPRC}. The results for $D$ are all found to be consistent with $\tilde D$ for all analyses based on $v^{p_i}$, {\it i.e.} a)-d). The paired-proton-cell approach, method e),  is known to be very sensitive to the proton-threshold effect discussed in section~\ref{sec:CoulterEffect} and produced a significantly different value. Correcting for the estimated proton-threshold effect on $D$ based on day-to-day correction of the $w^{p_ie_j}$  yielded a value consistent with $\tilde D$. Cross checks were also performed on data subsets and found to be consistent with the exception of the 31-kV data for which a large number of proton cells were dead.

A blind analysis of the asymmetries was adopted by adding a quantity $K_D^{p_i e_j}{\cal B}$ to each $w^{p_ie_j}$ (equation \ref{eq:ws}) so that
when $\tilde D$ was extracted from Eq.~\ref{eq:Dtilde} it was offset from the true value by ${\cal B}$, where $-0.01\le {\cal B} \le 0.01$. The factor ${\cal B}$ was revealed and subtracted as the final analysis step, after the corrections for systematic errors and all uncertainties were determined.

\subsection{Final Result}

When averaged over the entire run, each SBD was live for a majority of the time and had a high average efficiency. We can therefore combine counts for the entire run to determine the $v ^{p_i}$s and to extract $\tilde D$ for each set of 16 proton cells. The results for the four separate $\tilde D$ are presented in  Fig.~\ref{fg:DSuperSlices} and  Table~\ref{tb:DSuperSlices}. The weighted average of the four $\tilde D$ is $\langle\tilde D\rangle=(0.72\pm 1.89)\times 10^{-4}$. When combined with the total of corrections listed in Table~\ref{tb:Systematics}, our final result is
\begin{equation}
D=[-0.94 \pm 1.89 ({\rm stat}) \pm 0.97 ({\rm sys})]\times 10^{-4}.
\end{equation}
This differs from reference~\cite{rf:emiTPRL} due to refinement of the additive corrections due to background and backscattering, which changed by \mbox{$-0.07\times 10^{-4}$ and $0.09\times 10^{-4}$,} respectively.

\begin{figure}
\centerline{\includegraphics[width=3.25in]{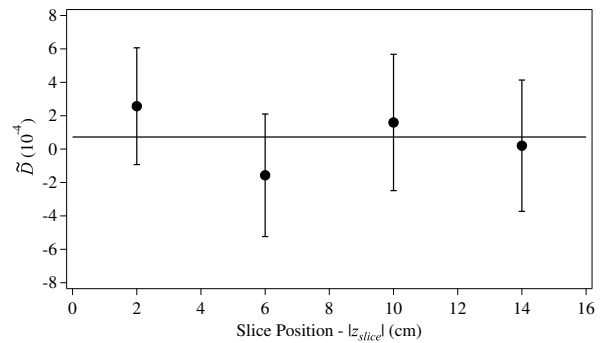}}
\caption{Results for $\tilde D$ for the entire experiment for each set of 16 proton cells. The  weighted average is $0.72\pm 1.89$ with $\chi^2=0.8$ for 3 degrees of freedom.}
\label{fg:DSuperSlices}
\end{figure}

\begin{table}[htdp]
 \caption{Results for the experiment for four proton-cell sets in units of $10^{-4}$. Uncertainties are statistical errors only.}
\begin{ruledtabular}
\begin{tabular}{ccc}
Proton-cell set & $|z|$ (cm)& \ \ $\tilde D$\\
1&2& $-2.57\pm 3.49$\\
2&6 &   $ -1.57  \pm  3.67  $\\
3&10 & $  \ \ 1.60 \pm  4.08$\\ 
4&14 & $\ \ 0.20 \pm  3.93$\\
Average &&$ \ \ 0.72  \pm  1.89$\\
\end{tabular}
\label{tb:DSuperSlices}
\end{ruledtabular}
\label{default}
\end{table}

\section{Conclusions}
\label{sec:Conclusion}

\begin{figure*}
\vskip-1 truein\includegraphics*[width=1\textwidth]{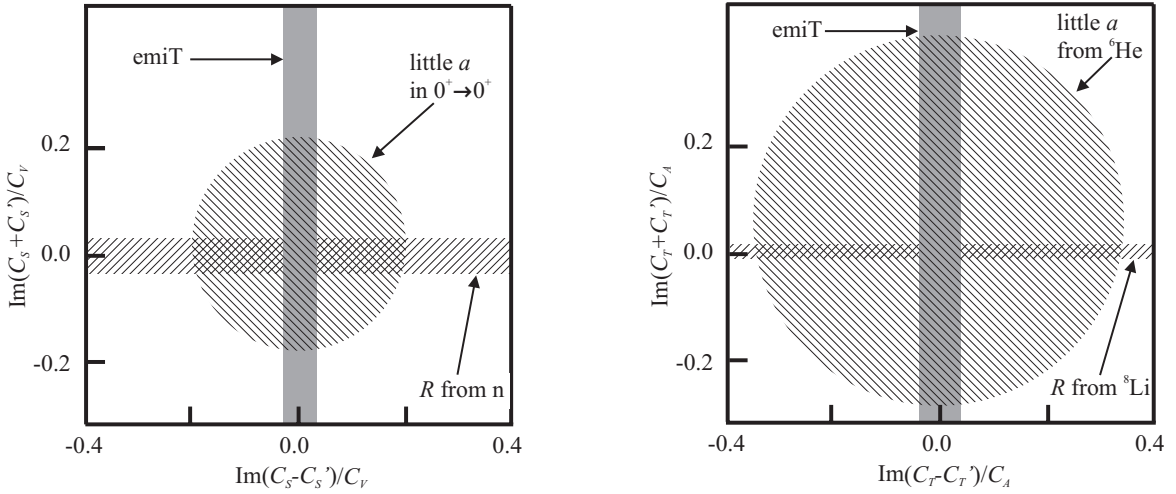}\vskip -1.25 truein
\caption{Sensitivity to $T$ violating couplings due to scalar and tensor currents from this work compared to limits from other experiments. Left: sensitivity to the imaginary part of  scalar couplings, assuming tensor currents to be purely real and equal to the largest presently-allowed value $\tilde C_T^-=0.12$. The bounds labeled ``$R~{\rm from}$ n'' are from Ref.~\protect\cite{rf:Kozela}, and those labeled ``little $a~{\rm in}~0^+\rightarrow 0^+$''  are from a combination of Ref.~\protect\cite{ad:99} and~\protect\cite{go:05}.
Right: sensitivity to the imaginary part of tensor couplings, assuming scalar currents to be purely real and equal to the largest presently-allowed value $\tilde C_S^- = 0.1$. 
The bounds labeled ``little $a$ from $^6$He'' are from reference~\protect\cite{Johnson:63} and  those labeled ``$R~{\rm from}~^8{\rm Li}$'' are from Ref.~\protect\cite{hu:03}. The current work is labeled ``emiT.'' All allowed regions represent 95\% confidence levels.}
\label{fig:scalartensor}
\end{figure*}

Our result represents the most sensitive measurement of $D$ in nuclear beta decay and can be interpreted in terms of possible extensions of the Standard Model. 
Rewriting Eq.~\ref{eq:DVAST}:
\begin{eqnarray}
D_{\not T}&=&\frac{2|\lambda|}{1+3 |\lambda|^2}\nonumber\\
&&\times \left[
\sin\phi_{AV}+
\frac{1}{2}{\rm Im}\left\{\tilde C_S^+ (\tilde C_T^{+ })^*+ \tilde C_S^- (\tilde C_T^{- })^*\right\}
\right],\nonumber\\
\end{eqnarray}
where $\tilde C_S^\pm=(C_S \pm C_S^\prime)/C_V$ and $\tilde C_T^\pm=(C_T\pm C_T^\prime )/C_A$, and
we have neglected the order-$\alpha$ radiative corrections, which yield negligible contributions given the existing limits on scalar and tensor currents.

Assuming no scalar or tensor currents, our result constrains the complex phase between the axial-vector and vector currents ($C_A/C_V=| \lambda |e^{i\phi_{AV}}$ ) to $\phi_{AV} = 180.012^\circ \pm 0.028^\circ$ (68\% confidence level).
If all currents are allowed, for example due tor leptoquark exchange, the equation contains five phases, making it difficult to compare the sensitivity of our experiment with respect to other probes without further assumptions. In the specific case where there are no special cancellations between terms, we estimate the sensitivity of our measurement under two different assumptions. Fig.~\ref{fig:scalartensor}, left panel, shows the limits on the imaginary component of the scalar currents, assuming that the tensor currents are purely real and equal to the largest value allowed by present constraints. In this case, because the limits on $\tilde C_T^+$ are much smaller than those on $\tilde C_T^-$ our result is most sensitive to the imaginary component of $\tilde C_S^-$.  The right panel shows a similar plot but now assuming that scalar currents are purely real and equal to the largest value allowed by present constraints.

\subsection{Potential Improvements}

 The result for $D$ presented here has comparable statistical and systematic uncertainties. Thus an improved experiment with the same apparatus would need both more neutron decays and reduced systematic effects. A new beam-line (NGC)  under construction at NCNR and the PF-1 beam at ILL could provide a factor of 10 or more decays per unit time. Reducing the three major systematic corrections requires 
eliminating the proton-threshold variations,
 a  more symmetric neutron beam, and smaller magnetic field. The symmetry of the neutron-beam is most strongly affected by the supermirror-bender neutron polarizer, while the 560~$\mu$T magnetic field was chosen to effect sufficient velocity averaging of transverse-neutron polarization produced in the current-sheet spin flipper. An alternative polarizer is a steady-state polarized $^{3}$He spin filter~\cite{rf:ChuppNIM}. Intense cold neutron beams have been shown to affect the rubidium and $^3$He polarization~\cite{rf:Sharma2008}; however, this appears to be a solvable technical challenge~\cite{rf:Babcock}. The 560~$\mu$T guide field can be reduced by using an adiabatic-fast-passage neutron spin flipper and effective shimming of the magnetic field along with shielding of external field perturbations. Thus a factor of three or more improvement in sensitivity  to $D$ appears within reach with the current apparatus. Extending the sensitivity to the level of final-state-effects ($D_{FSI}\approx 10^{-5}$) and beyond is a well motivated goal that would  require an apparatus with  greater geometric efficiency for both proton and electron detectors.

\section*{Acknowledgements}
The authors gratefully acknowledge informative conversations with Wick Haxton, Michael Ramsey-Musolf, and Sean Tulin. The emiT apparatus and emiT-I experiment were developed by many additional collaborators including Shenq-Rong Hwang, Laura Lising, Hamish Robertson, and Tom Steiger  along with Bill Teasdale.
The neutron facilities used in this  work were provided by the National Institute of Standards and Technology, U.S. Department of Commerce. The research was made possible in part by grants from the U.S. Department of Energy Office of Nuclear Physics (DE-FG02-97ER41020, DE-AC02-05CH11231, and DE-FG02-97ER41041), and the National Science Foundation (PHY-0555432, PHY-0855694, PHY- 0555474, and PHY-0855310).

\renewcommand*{\refname}{}

\end{document}